\documentclass[11pt,a4paper]{article}
\pdfoutput=1
\usepackage{jheppub}
\usepackage[dvipsnames]{xcolor}
\usepackage{colortbl}
\usepackage{tabularx}
\usepackage{arydshln}
\usepackage{bbm}
\usepackage{subcaption}
\usepackage{tkz-fct}
\usepackage{pgfplots}

\pgfplotsset{compat=1.18}

\usetikzlibrary{backgrounds}
\usepgfplotslibrary{fillbetween}

\usepackage[dvipsnames]{xcolor}
\usepackage{amsthm}
\usepackage{amsmath}
\usepackage{amssymb}
\usepackage{mathtools}
\usepackage{dsfont}
\usepackage{graphicx}
\usepackage{multirow}
\usepackage{scalerel}
\usepackage{tikz}
\usetikzlibrary{calc,decorations.markings}
\usepackage{esint}
\usepackage{soul}
\usepackage{appendix}
\usepackage{hyperref}

\renewcommand*{\le}{\left}
\newcommand*{\ri}{\right}

\renewcommand*{\b}{\beta}
\newcommand*{\g}{\gamma}
\renewcommand*{\d}{\delta}
\newcommand*{\e}{\epsilon}

\renewcommand*{\l}{\lambda}

\renewcommand*{\r}{\rho}

\newcommand*{\f}{\phi}

\renewcommand*{\k}{\kappa}
\newcommand*{\z}{\zeta}

\newcommand*{\G}{\Gamma}

\newcommand*{\p}{\partial}

\newcommand*{\diff}{\mathrm{d}} 
\newcommand{\mi}{\mathrm{i}} 
\newcommand{\rme}{\mathrm{e}}
\newcommand{\sph}[1][]{\ifmmode \mathrm{S}^{#1} \else S\(^{#1}\)\fi} 

\newcommand*{\cB}{\mathcal{B}}

\newcommand*{\cI}{\mathcal{I}}

\newcommand*{\N}{\mathcal{N}}
\newcommand*{\cO}{\mathcal{O}}
\newcommand*{\cZ}{\mathcal{Z}}
\newcommand*{\beq}{\begin{equation}}
\newcommand*{\eeq}{\end{equation}}
\newcommand*{\bea}{\begin{eqnarray}}
\newcommand*{\eea}{\end{eqnarray}}

\newcommand*{\fm}{\mathfrak{m}}
\newcommand*{\fa}{\mathfrak{a}}

\newcommand*{\gym}{g_{\text{YM}}}
\newcommand*{\Z}{\mathcal{Z}_{\text{ScYM}}}

\definecolor{light_gray}{HTML}{d4d4d4}
\definecolor{lighter_gray}{HTML}{dcdedc}
\definecolor{light_red}{HTML}{eda29a}

\hypersetup{
    colorlinks=true,
    linkcolor=blue,
    filecolor=magenta,      
    urlcolor=blue,
    citecolor=blue,
    pdftitle={Overleaf Example},
    pdfpagemode=UseOutlines,
}

\title{The resurgence of errors in the localization of $\N = 2$ superconformal Yang-Mills}

\author[1,3]{Inês Aniceto}
\author[2,3]{James Ratcliffe}
\author[1,3]{Itamar Yaakov}

\affiliation[1]{School of Mathematical Sciences, University of Southampton,
Highfield, \\
    Southampton SO17 1BJ, United Kingdom}
\affiliation[2]{Department of Physics and Astronomy, University of Southampton, Highfield, \\
Southampton SO17 1BJ, United Kingdom}
\affiliation[3]{STAG Research Centre, University of Southampton, Highfield, \\
Southampton SO17 1BJ, United Kingdom}

\emailAdd{I.Aniceto@soton.ac.uk}
\emailAdd{J.Ratcliffe@soton.ac.uk}
\emailAdd{I.Yaakov@soton.ac.uk}

\abstract{ 
We give a physical interpretation for the analytic continuation of the partition function of superconformal SU$\left(2\right)$ $\mathcal{N}=2$ gauge theory on the four-sphere to all values of the Yang-Mills coupling. We show that a well-motivated 2d construction associates two-dimensional unstable instantons to the 4d complex saddles which appear as singularities in the integrand of the supersymmetric localization expression. The construction is based on the chiral algebra subsector, and aligns with the alternative Higgs branch localization.}

\begin{document}

\maketitle

\section{Introduction}

Non-perturbative effects in QFT are usually associated with very visible, stable, and often topologically protected, classical solutions of the equations of motion. Although these are the most straightforward configurations to identify, others can be equally important in addressing specific physical questions. 
Resurgent analysis of path integrals offers a holistic approach to the study of the analytic structure of coupling dependence of QFT observables, incorporating both perturbative and non-perturbative effects, and famously linking the two (see \cite{sauzin_lecture_notes, Aniceto_primer} for detailed reviews). In fact, resurgence is a natural framework with which to study all non-perturbative effects via analytic continuation, often picking up terms overlooked by a ``physical'' regime. In this paper, we study a situation in which neither standard perturbation theory nor topologically protected classical solutions offer a good guide to the physics involved. Instead, we will use resurgent analysis to study somewhat more subtle non-perturbative contributions to an observable, eventually identifying them with unstable configurations, with the help of supersymmetric localization.

The context for our analysis is the exact expression for supersymmetric observables of 4d $\mathcal{N}=2$ gauge theories. We will concentrate on the partition function of the theory coupled to curvature on the round four-sphere in a supersymmetric manner. The path integral for this observable was shown to reduce, via supersymmetric localization, to a finite-dimensional integral over a bosonic degree of freedom valued in the Cartan subalgebra for the gauge group \cite{Pestun:2007rz}, referred to as the Coulomb-branch. The integrand, or 1-loop determinant, comprising ratios of multiple gamma functions, is the result of applying an exact saddle point approximation. For gauge group SU$(2)$, the salient piece of this integrand is given by
\beq\label{eq:Z_1_loop_SU(2)}
    Z^{\text{SU}(2)}_{1-\text{loop}}(a) = \prod_{n = 1}^\infty\frac{\le( 1 + \frac{4a^2}{n^2}\ri)^{2n}}{\le(1 + \frac{a^2}{n^2} \ri)^{8n}}\,.
\eeq
Additionally, the expression includes a sum over non-perturbative contributions associated with supersymmetric instantons and anti-instantons, which also contribute to the integrand. These are classical configurations, of the visible type alluded to above, which will not play a role in our analysis. Instead, we will provide a physical interpretation of the poles in eq. \eqref{eq:Z_1_loop_SU(2)} in terms classical configurations of an associated 2d quantum field theory. We will motivate the action and symmetries of this theory using the chiral algebra construction of \cite{Beem_2015}, and specifically its incarnation on the four-sphere studied in \cite{Pan:2017zie}.

The integrals defining the Coloumb-branch expressions are convergent, but the integrands have poles in the complex plane. From a purely mathematical perspective, resurgent analysis of these integrals reduces to the residue theorem. The utility of the holistic resurgence approach in this case is in helping to identify the physical meaning of the poles, rather than as a tool for computation. As we will show, summing over the poles does not make sense for physical values of the Yang-Mills coupling, and the correct interpretation is that of analytic continuation of the partition function to complex values of the coupling. An analysis of the poles, treating the expression as a Laplace integral in the Borel plane, will also allow us to make a well-informed hypothesis as to their physical origin. We note that the type of Borel plane singularities we encounter resembles a simplified version of the Peacock pattern observed in the context of topological string theory \cite{Garoufalidis:2020xec,Fantini:2024snx}.

The study of singularities in the Borel plane is essential for understanding the non-perturbative effects which contribute to the analytic expression for  QFT or string-theory observables, and which would otherwise be missed when restricting to certain ``physical settings''. For instance, in quantum mechanics, tunnelling effects can be obtained by analytically continuing coupling constants to negative values \cite{Bender:1969si,BenderWu1973}. In QFTs, there exist \textit{renormalons} - non-perturbative contributions from loop-integrals in perturbation theory contributing to their $n!$ growth. Their presence in the UV is signalled by singularities in the Borel plane along the negative real axis \cite{Parisi1978,Parisi1979,Lautrup1977, tHooft1979}. Other cases of additional configurations predicted by resurgent analysis, in QFT and string theoretic observables, can be found in \textit{e.g.} \cite{Cherman:2013yfa} (fractons) and \cite{Marino:2022rpz,Schiappa:2023ned} (negative tension D-branes). In the context of localizable observables, the role of Borel plane singularities and resummation procedures was studied in \cite{Aniceto:2014hoa,Honda:2016vmv,Gukov:2016njj,Honda:2017cnz,Dorigoni:2017smz,Gang:2017hbs,Fujimori:2021oqg,Dorigoni:2021guq}. 

Depending on their matter content, some localizable supersymmetric models admit multiple localization schemes. Of particular interest to the present work are ``Higgs branch'' and ``Coulomb branch'' localization schemes. Since localization is independent of how one chooses to localize, the results computed in both schemes should be equal. In fact, Higgs-branch modes often appear as singularities in Coulomb-branch computations. The origin of these singularities is well understood, for some models, in terms of vortex and anti-vortex field configurations \cite{Benini_2014_2D,Benini_2014,Bullimore_2019,bullimore2023twistedindextopologicalsaddles, Fujitsuka_2014,Peelaers_2014}. In four-dimensional theories on $\mathbb{S}^4$, the authors of \cite{PanPeelaers} were able to produce a Higgs-branch-localized expression, but the connection to the Coulomb-branch there is less clear. In this context, following the work of \cite{PanPeelaers}, we will elaborate on the way in which the Higgs-branch expression is related to singularities of the Coulomb-branch integral, eventually connecting this to our two-dimensional model.
\newline
\newline
\textbf{Summary of results:} We begin by discussing the physical importance of studying non-perturbative objects in complex $\gym$ space and introduce the physical model in four-dimensions in sec. \ref{sec:Higgs}. We will obtain a Higgs-branch form of the 4d path integral by identifying monopole configurations that become important at imaginary values of $\gym$, showing that it is analytically related to the Coulomb-branch form of the partition function. In sec. \ref{sec:Resurgence_Higgs}, we use resurgent analysis to show that the singularities in the Coulomb branch are described by the non-perturbative data of a sum of error functions, strongly implying a connection to two-dimensional physics. In sec. \ref{sec:pole_physics}, we make contact with the chiral algebra subsector, and reproduce the four-dimensional expression using 2d supersymmetric localization. We conclude and summarize our results in sec. \ref{sec:discussion} and discuss how they can be generalized to a wide class of superconformal models in four-dimensions.

\section{Higgs branch without localization} \label{sec:Higgs}
In this work, we will be interested in $\N = 2$, SU$(N_c)$
super-Yang-Mills (SYM) theory with $N_f$ fundamental massless hypermultiplets coupled in a supersymmetric manner to curvature on the four sphere. Following \cite{Pestun:2007rz}, the partition function for this theory localizes to a gauge-fixed matrix model given by the integral\footnote{\label{fn:prop} The proportionality symbol represents a numerical factor outside the integral that we will later introduce. This normalizes the integral such that it equals $1$ in the free theory limit, $\gym \to 0$. This is not the usual normalization for the partition function, but it is the most convenient one for our purposes.}
\beq \label{eq:Pestun_MM}
    \cZ \propto \int \mathrm{d}^{N_c-1} \l \prod_{1 \leq i < j \leq N_c} \le(\l_i-\l_j\ri)^{2} \rme^{-\frac{2}{\gym^{2}}\sum_{i=1}^{N_c}\l_i^2} Z_{1 \mathrm{-loop}}(\l)\le| Z_{\mathrm{inst}}\left(\gym,\lambda\right)\ri|^2\,,
\eeq
where $\l_i$ parametrize the Cartan subalgebra, and $Z_{\text{inst}}$ is the Nekrasov instanton partition function \cite{Nekrasov:2002qd}. We refer to this as the Coulomb-branch expression for the partition function. 

Consider the simplest case of SU$(2)$ gauge group. For this choice, the matrix model \eqref{eq:Pestun_MM} reduces to
\beq \label{eq:Z_SU(2)}
    \cZ = \frac{128 \pi^{5/2}}{\gym^3} \int_{-\infty}^{\infty} \diff a \, a^2 \,  \rme^{-\frac{16 \pi^2}{\gym^2}a^2} \frac{(G(1 + 2a \mathrm{i})G(1 - 2a \mathrm{i}))^2}{\phantom{{}^{N_f}}(G(1 + a \mathrm{i})G(1 - a \mathrm{i}))^{2 N_f}} \, |Z_{\text{inst.}}|^2\,,
\eeq
where we have introduced the normalization constant mentioned in footnote \ref{fn:prop}, and where $G(1 + z)$ is the Barnes--G function,
\beq \label{eq:Barnes_G_prod}
    G(1 + z) = (2\pi)^{z/2} \, \rme^{-((1 + \g z^2) + z)/2} \prod_{n = 1}^\infty \le( 1 + \frac{z}{n} \ri)^n e^{-z + \frac{z^2}{2n}}\,.
\eeq
We will introduce the following shorthand for the combination of Barnes--G functions seen above
\beq \label{eq:Bernes-comb}
    \mathcal{G}_{N_f}(a) \equiv \frac{(G(1 + 2 a \mathrm{i})G(1 - 2 a \mathrm{i}))^2}{\phantom{{}^{N_f}}(G(1 + a \mathrm{i})G(1 - a \mathrm{i}))^{2 N_f}}\,.
\eeq
As detailed in \cite{Aniceto:2014hoa}, using the product representation of the Barnes--G function, \eqref{eq:Barnes_G_prod}, it can be seen that the singularities of the integrand are poles located along the imaginary $a$ axis. Similarly, the integral expression for the Coulomb branch localization of an SU$(N_c)$  theory will have an integrand with only pole singularities, and these will lie away from the real axis.   

There also exists a Higgs branch expression for the partition function of the SU$(2)$ theory, which can be found in \cite{PanPeelaers}.\footnote{The analysis in \cite{PanPeelaers} is for U$(N)$ theories, and employs an FI parameter which is not available for SU$(N)$. The authors use this parameter to suppress certain contributions to the path integral. We believe most of the relevant results are still valid, in the superconformal case, without this parameter. Specifically, the BPS equations and the configurations which solve them are still valid.} The physical configurations in the Higgs branch localization scheme are partially indexed by an integer $n$ which captures the winding number of smooth vortex-like configurations and Chern classes of singular Seiberg-Witten monopoles.\footnote{We are working in the round four-sphere limit, i.e. with vanishing squashing parameter, where the two integers which classify the BPS configurations in \cite{PanPeelaers}  degenerate to just one.} The classical contribution for such configurations is given as

\beq \label{eq:Higgs_modes}
\mathrm{e}^{-S_\text{classical}}=\exp\left(\frac{16\pi^2 n^2}{\gym^2}\right)\,.
\eeq

\noindent
Comparing with the Coulomb branch expression (\ref{eq:Z_SU(2)}), we will see that we can identify the exponentiated action of these configurations as part of the \textit{residue} at the poles for the Coulomb branch integrand. This is a typical scenario when relating the expressions for the two schemes. We therefore tentatively identify the two, and attempt to determine the contribution of these configurations through a residue calculation of the Coulomb branch integrals. 

Our goal, therefore, is to express the partition function $\cZ$ of \eqref{eq:Z_SU(2)} purely in terms of Higgs branch objects, and to properly understand their role. It is clear from \eqref{eq:Higgs_modes} that these objects cannot participate for real $\gym$ without resummation, as they would be exponentially enhanced. Therefore, the expression obtained from simply summing over these terms will have a proper subregion of the complex plane as its domain of convergence. We must therefore determine under what conditions can $\cZ$ be written as a sum over Higgs-branch objects, or equivalently, as a sum over residues. To answer this question, we will analyze the large--$|a|$ behavior of the integrand in $\cZ$. 
 
 Asymptotically, as $|z| \to \infty$ for $|$arg$(z)| < \pi$,\footnote{Our phase convention is such that $\text{arg}(z)$ runs from $-\pi$ to $\pi$.} the Barnes--G function behaves as \cite{NIST:DLMF}
\beq
G( 1 + z ) \approx \text{exp} \le( \frac{3}{4}z^2 + \le( \frac{z^2}{2} - \frac{1}{12} \ri)\text{log}(z) + \frac{z}{2}\text{log}(2\pi) \ri)\,.
\eeq
The large--$|a|$ behaviour of the full quotient $\mathcal{G}_{N_f}(a)$ in \eqref{eq:Bernes-comb} is then
\beq \label{eq:1loop_asym}
	\mathcal{G}_{N_f}(a) \approx \text{exp} \le( (N_f - 2N_c) ((2a^2 + 1/3)\text{log}(a) -3a^2)- \le(4 a^2 \ri)N_c \, \text{log}(2) \ri)\,.
\eeq
The Vandermonde determinant contributes as log$(a)$ and will therefore not contribute to our analysis. The large--$|a|$ properties of the integrand in \eqref{eq:Z_SU(2)} are therefore given by a competition between \eqref{eq:1loop_asym} and the $\gym$--dependent contribution
\beq \label{eq:integrand_large_a_behaviour}
    \rme^{-\frac{16 \pi^2}{\gym^2}a^2}\mathcal{G}_{N_f}(a) \approx\text{exp}\le( - \frac{16 \pi^2}{\gym^2}a^2 + (N_f - 2N_c) (2a^2\text{log}(a) -3a^2)- 4 a^2 N_c \, \text{log}(2)\ri).
\eeq
There are three cases to consider. If $N_f > 2 N_c$, then the integral expression in \eqref{eq:Z_SU(2)} diverges and is not well defined. Such theories are not asymptotically free. Conversely, if $N_f < 2 N_c$, then \eqref{eq:integrand_large_a_behaviour}, evaluated at a pole $a = \mi k$, behaves for large $k$ as
\beq
    \text{exp}\le(\frac{16\pi^2}{\gym^2}k^2 \ri)\mathcal{G}_{N_f}(\mi k) \approx \text{exp}\le( 2 k^2 (2N_c - N_f)\,\text{log}(k) \ri).
\eeq
Therefore, the \textit{sum} over all residues diverges for any value of $\gym$. The final case is $N_f = 2 N_c$ where the problematic term in \eqref{eq:integrand_large_a_behaviour} vanishes. Recalling that the $\beta$--function of $\N = 2$ theories with $N_f$ hypermultiplets behaves as
\beq
	\b_{\N = 2} \sim - 2 N_c + N_f,
\eeq
we see that for this matter content the theory is superconformal. We conclude that the original integral \eqref{eq:Z_SU(2)} and the sum over the residues are simultaneously meaningful, possibly for different values of $\gym$, \textit{only when the theory is superconformal}. We will henceforth assume this criterion and refer to the theory as Superconformal Yang-Mills (ScYM), denoting the partition function as $\Z(\gym)$.

We can now proceed and determine the exact form of the residues. We will conduct our analysis within the zeroth instanton sector only. By this, we strictly mean the sector without contributions from instantons or anti-instantons, as opposed to the sector with zero instanton charge. The analysis for higher-order instanton sectors follows by analogy: $Z_{\text{inst}}$ contributes poles along the imaginary axis of $a$ of order strictly less than the zeros of $\mathcal{G}_{N_f}(a)$ and at different positions, as noted in \cite{Aniceto:2014hoa}. Thus, the singularity structure is the same in every instanton sector, but the residues will be altered.

For $N_f = 2 N_c = 4$, we find that $\mathcal{G}_4(a)$ has poles at $a = \pm \mathrm{i} \, n$, $n \in \mathbb{Z}_+$ of order $4n$. To compute the residues, we follow \cite{Aniceto:2014hoa}, and take $4n - 1$ derivatives of 
\beq \label{eq:I(a)}
    \mathcal{J}(a)\coloneqq(a - a_n)^{4n} \mathcal{I}(a),
\eeq
and take the limit $a \to a_n$, where $\mathcal{I}(a)$ is the integrand defined by taking $N_f=2N_c=4$ in zeroth instanton sector of \eqref{eq:Z_SU(2)}
\beq \label{eq:Z_ScYM}
    \Z^{k=0}(\gym,0) \coloneqq \frac{1}{\gym^3} \int_{-\infty}^\infty \diff a \, \mathcal{I}(a).
\eeq
The meaning of the second argument above will be introduced shortly. We will, in general, drop the subscript $k=0$ and implicitly assume that we are working in the zeroth instanton sector, in the manner described above. The derivation of the residues is carried out in appendix \ref{app:Residues}. The explicit expressions for the Laurent series of $\cI(a_n)$ are lengthy, involving polynomials in values of the Riemann $\z$-function. For our purposes, it is sufficient to note that the residues take the schematic form
\beq \label{eq:residues}
    \mathcal{R}(n) := \frac{1}{\gym^3}\text{Res}[\mathcal{I}(\pm\, \mi n)] = \frac{1}{2\gym^3} \text{exp} \le( \frac{16 \pi^2 n^2}{\gym} \ri)\sum_{\ell = 0}^{4n - 1} \frac{(-1)^{4n - \ell - 1}}{\G(4n - \ell)\ell!} \, f_{n,\ell}\, \gym^{-2(4n-1-\ell)}\,,
\eeq
where $f_{n,\ell}$ are the coefficients of a Borel transform associated to $\Z$, to be introduced in sec. \ref{sec:Resurgence_Higgs}. They are related to coefficients, $\hat{f}_{n,\ell}$, in the Laurent series of $\cI(a)$ without the exponential
\beq \label{eq:series_I_no_exp}
    128 \pi^{5/2}a^2 \frac{(G(1 + 2a \mi)G(1 - 2a \mi))^2}{(G(1 + a \mi)G(1 - a \mi))^8} \sim \frac{1}{(a - \mi n)^{4n}}\sum_{\ell = 0}^{\infty}\hat{f}_{n,\ell}\,(a - \mi n)^\ell\,.
\eeq
All these coefficients are derived and displayed in appendix \ref{app:Residues}. Presently, however, we are interested in computing the sum over the residues, \eqref{eq:residues}, and so we would like to understand its convergence properties, governed by the behaviour of the residues as $n \to \infty$. This large--$n$ behaviour is
\beq
\begin{aligned}
    \mathcal{R}(n \to \infty) \sim & \, \text{exp} \le[\frac{16\pi^2n^2}{g_{\text{YM}}^2}+\sum_{k=1}^{n}(12k-8n)\text{log}(k)-\sum_{k=n+1}^{2n}(4k-8n)\text{log}(k)\ri]
    \\ 
    \sim & \, \text{exp} \le[\frac{16\pi^2n^2}{g_{\text{YM}}^2}+8n^2 \text{log}(2) \ri]\,.
\end{aligned}
\eeq

Comparing this to the large--$a$ behavior of the integrand $\mathcal{I}(a)$ in \eqref{eq:Z_ScYM},
\beq \label{eq:asym_Z}
\begin{aligned}
    \mathcal{I}(a) & \simeq \text{exp} \le[ - \le(\frac{16\pi^2}{\gym^2} a^2 + 8a^2 \text{log}(2) \ri) \ri]\,,
\end{aligned}
\eeq
we see that, for real $a$, the regions of convergence in $\gym$ space of the integral and infinite sum of residues are complementary, \textit{i.e.} when one converges, the other diverges. This suggests that if there exists a relationship between the original integral and the sum over residues, it must be through analytic continuation.
 
\begin{figure}
    \centering
    \includegraphics[trim=1 5 1 5]{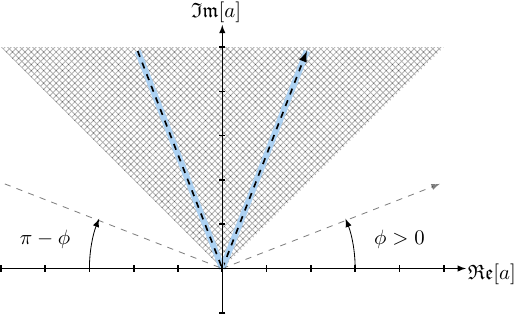}
    \caption{Pictorial representation of rotating the contour of integration $a \in (-\infty, \infty)$. For contours within the shaded region, $\pi/4 < \text{arg}(a) < 3\pi/4$, $\gym$ can be chosen such that these contours are equal to the sum over residues. An example of such a contour is highlighted in blue.}
    \label{fig:a_continuation}
\end{figure}

\begin{figure}
    \centering
    %
    %
    \begin{subfigure}{0.45\textwidth}
    \centering

    \includegraphics[width=\textwidth]{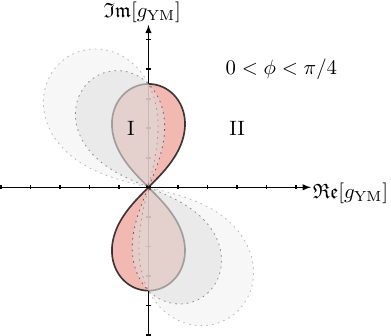}
    \caption{}
    \label{subfig:figures_of_8}
    
    \end{subfigure}
    \hspace{2.5em}
    %
    %
    \begin{subfigure}{0.45\textwidth}
    \centering
    
    \includegraphics[width=\textwidth]{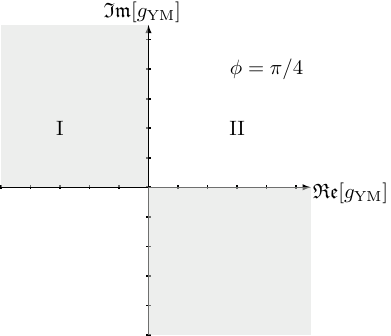}
    \caption{}
    \label{subfig:infinite_8}
    
    \end{subfigure}\\
    %
    %
    \begin{subfigure}{0.45\textwidth}
    \centering

    \includegraphics[width=\textwidth]{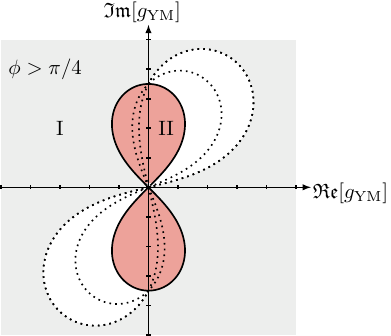}
    \caption{}
    \label{subfig:inverted_8}
    
    \end{subfigure}
    
    \caption{Diagrams depicting the convergent behaviour of the positive contour defining $\cZ(\gym,\f)$. This corresponding rotation in the $a$--plane is shown in figure \ref{fig:a_continuation}. In \ref{subfig:figures_of_8} we rotate the contour by $\f$, and see the divergent region rotate anti-clockwise and expand. The limiting case is shown in \ref{subfig:infinite_8}, when $\f = \pi/4$: the divergent region is the entire 2nd and 4th quarter planes. If we continue to rotate arg$(a) > \pi/4$ the domains flip, and inside the figures of 8 is where the integral converges. The shaded red region is where the formal sum of residues converges. The analogous behaviour for the negative contour defining $\Z(\gym,\f)$ is given by reflecting these diagrams about the $y$--axis.}
    \label{fig:figures_of_8}
\end{figure}

Let us explain this in more detail. Although the domains of convergence of $\Z$ and the sum over residues are disjoint, the contour defining $\Z$ can be rotated so that the rotated domain of convergence overlaps with both the sum over residues and the original contour. In this sense, the residues and original contour are analytically related. To make this concrete, we will consider a one-parameter family of contours: where $a\in (-\infty,0)$ and $a \in (0,\infty)$ are rotated into the upper-half complex $a$-plane by a phase $\f$, as depicted in fig. \ref{fig:a_continuation}. We will refer to these contours as the negative and positive contours, respectively. We denote the corresponding partition function $\Z(\gym,\f)$. The original contour corresponds to $\f = 0$, where $\Z(\gym,0)$ was defined earlier in eq. \eqref{eq:Z_ScYM}. From the large-$a$ behavior of the integrand \eqref{eq:asym_Z}, we conclude that, for $\f < \pi/4$, so that arg$(a)<\pi/4$ or arg$(a)>3\pi/4$, the integral expression for $\Z(\gym,\f)$ converges when
\beq
    \mathfrak{Re}\le[ \frac{a^2}{\gym^2} \ri] \geq \mathfrak{Re} \le[- \frac{8 \text{log}(2)}{16 \pi^2} a^2\ri]\,.
\eeq
For $\f>\pi/4$, the inequality changes direction. Explicitly, taking arg$(\gym) = \varphi$ and $\f>\pi/4$ we find 
\beq
    \frac{2 \pi^2}{\text{log}(2)} \frac{\text{cos}(2(\varphi + \phi ))}{ \text{cos}(2\phi)} + |\gym|^2 \geq 0\,.
\eeq
For a given value of $\phi$, this carves out the entire complex $\gym$ plane, except for a region enclosed by a figure-of-eight, outside of which $\Z(\gym,\f)$ converges. As we increase $\f$, the figure-of-eight rotates (anti-clockwise) clockwise and grows for the (positive) negative contours. This behaviour, for the positive contour, is shown in fig. \ref{fig:figures_of_8}. The divergent regions are denoted by I, while the convergent regions are denoted by II. The sum over residues is instead convergent within a fixed figure of eight, depicted by the pink-shaded region in fig. \ref{subfig:figures_of_8}. We will denote the interior of this figure-of-eight as $\mathfrak{D}$. 
For the rotated contour integral to equal the sum over residues, we require both halves of the contour, and the residue sum, to converge simultaneously for a given value of $\gym \in \mathfrak{D}$. 

We can determine the appropriate contours as follows. As we rotate, the convergent regions of each half-contour begin to overlap with the sum of residues. However, they do not yet mutually overlap. Eventually, a critical point is reached when $\f = \pi/4$, where the divergent region fills the entire 2nd and 4th quadrants of the complex $\gym$--plane for the positive cofntour, as shown in fig. \ref{subfig:infinite_8}, and fills the 1st and 3rd quadrants for the negative contour. Continuing past this point, the divergent region I and convergent region II swap: the integral converges within an area enclosed by the figures of eight shown in fig. \ref{subfig:inverted_8}. Moreover, the convergent regions of both half-contours mutually overlap with the convergent region of the residue sum: both half contours in this region converge and are equal to the sum over residues for $\gym \in \mathfrak{D}$. In the complex $a$--plane this region is the gray-shaded region in fig. \ref{fig:a_continuation} and an example of such a contour is highlighted blue. 

From the diagrams in fig. \ref{fig:figures_of_8} we can also see that the domain of convergence when $\f = \pi/4$, depicted in fig. \ref{subfig:infinite_8}, overlaps with both the convergent region of the original contour in fig. \ref{subfig:figures_of_8} and the convergent region of the sum over residues, as mentioned above. We have confirmed this convergence and equality between the integral and the sum explicitly to third order in the residue sum for various values of $\gym$ in the domain of convergence. For example, for $\gym = 4 \mi$ we have
\beq
\begin{aligned}
    \Z(4 \mi) &= 3.98817 \, \mi\,;
    \\
    \Z(4 \mi) - 2 \pi i \,\mathcal{R}(1) & = 0.00199463 \, \mi \,;
    \\
    \Z(4 \mi) - 2 \pi i \, (\mathcal{R}(1) + \mathcal{R}(2)) & = -2.89333 \times 10^{-11} \, \mi\,,
\end{aligned}
\eeq
where the residue $\mathcal{R}(n)$ is defined in \eqref{eq:residues} with coefficients given in appendix \ref{app:Residues}. We note that the final equality is of the order of the third residue, $2 \pi \mi \, \mathcal{R}(3) =-2.93979 \times 10^{-11}$.

We have belaboured this point for good reason. It is not true that the path integral is generically equal to the sum over residues, as Higgs-branch localization usually prescribes. Rather, \textit{it is equal through analytic continuation}. The path integral $\Z(\gym)$ is independent of $\f$ and is an analytic function of the coupling. The integral and sum should be interpreted as \textit{representations} of this function, each valid within a certain domain. The domains where one formal expression is valid while the other is not should be considered as analytic continuations of each other. The full function $\Z (\gym)$, if such a quantity could be computed explicitly, should be viewed as the complete resummation of these partial pictures. In the ideal case, this result can be re-summed into a form where the analytic properties of $\Z$ with respect to $\gym$ are immediate. Recall, for instance, the equality $\sum_n x^n = 1/(1-x)$. Here, the sum is a representation of the analytic function, valid when $|x| < 1$. The function $1/(1 - x)$ is thus the analytic continuation of the geometric series. In our case, however, explicit resummation is a difficult task due to the complicated form of the coefficients $f_{n,\ell}$ of each residue \eqref{eq:residues}, and the \textit{quadratic} dependence on $n$ in the exponential factor. 

Our final result for the four-dimensional partition function is thus
\beq
\left.
\begin{aligned}
    \Z^{k=0}(\gym) &= 2\pi \mi\sum_{n = 0} \mathcal{R}(n)\,
    \\
    &= \frac{\pi \mi}{\gym^3}\sum_{n = 0} \text{exp} \le( \frac{16 \pi^2 n^2}{\gym} \ri)\sum_{\ell = 0}^{4n - 1} \frac{(-1)^{4n - \ell - 1}}{\G(4n - \ell)\ell!} \, f_{n,\ell}\, \gym^{-2(4n-1-\ell)}
\end{aligned}
\right\}
\quad
\gym \in \mathfrak{D}\,.
\eeq

To conclude, from the form of the residues \eqref{eq:residues} we can interpret the exponential terms as classical actions of Seiberg-Witten monopoles, by comparing to the localization results of \cite{PanPeelaers}. This matches the expectation that this result can be considered as the Higgs-branch form of the Coulomb-branch integral. It would be interesting to explicitly compute the Higgs-branch localization of this theory to see how it compares with the sum of residues: is it exactly the result we have obtained or an analytic continuation of it? We will not investigate this question in this work. Instead, we would like to understand what the Higgs-branch result \eqref{eq:residues} corresponds to physically, in the region in which it converges, and interpret the Coulomb-branch form of $\Z$ as its analytic continuation. To this end, in the following section we will start from the results of \cite{Russo_2012,Aniceto:2014hoa} and employ resurgent analysis to gain some insight into the meaning of these complex saddles.

\section{Localization is full of errors} \label{sec:Resurgence_Higgs}
We would like to provide a physical interpretation of the residues \eqref{eq:residues}. To do so, we build on the results of \cite{Russo_2012, Aniceto:2014hoa}, by identifying the original integral \eqref{eq:Z_SU(2)} with a sum of Laplace transforms, thus allowing resurgent analysis to be applied. We first review the relevant results of \cite{Aniceto:2014hoa}. There, it was shown that the separate halves of the Coulomb branch integral have a divergent asymptotic series in $\gym^2$. We can show this explicitly as follows. Splitting the contour of integration of $\Z(\gym,\f)$ into the positive and negative halves we introduced earlier, we have, explicitly,
\beq \label{eq:integral_splitting}
\begin{aligned}
    \Z(\gym,\f) = &~ \frac{128 \pi^{5/2}}{\gym^3} \le[ \int_{\rme^{\mi(\pi- \f)}\infty}^0 \diff a_-\; a_-^2\; \text{exp}\left(\frac{-16\pi^2}{g^2_{\text{YM}}}a_-^2\right) \mathcal{G}_4(a_-) \ri.
    \\    
    & \phantom{128\pi^{5/}-} +\le. \int^{+\rme^{\mi \f}\infty}_0 \diff a_+\; a_+^2\; \text{exp}\left(\frac{-16\pi^2}{g^2_{\text{YM}}}a_+^2\right) \mathcal{G}_4(a_+) \ri]
    \\
    = & ~ \frac{1}{\gym^3}\le(\mathcal{W}(\gym,\f) - \mathcal{W}(\gym,\pi-\f)\ri)\,,
\end{aligned}
\eeq
where $\mathcal{G}_4(a)$ is given in \eqref{eq:Bernes-comb}, and we have introduced the shorthand notation for the half integrals
\beq 
\begin{aligned}
    \mathcal{W}(\gym,\f) &\coloneqq 128 \pi^{5/2}\int_0^{+\rme^{\mi \f}\infty} \diff a \, a^2 \, \text{exp} \left(\frac{-16\pi^2}{g^2_{\text{YM}}}a^2\right) \mathcal{G}_4(a)\,.
\end{aligned}
\eeq

Each integral $\mathcal{W}$ can be rewritten as a Laplace-transform via the function $a = \frac{\sqrt{s}}{4\pi}$. The result of this map on $\mathcal{W}$ is
\beq \label{eq:W}
    \mathcal{S}_\theta \mathcal{W}(\gym, \theta) = \frac{1}{\sqrt{\pi}}\int_0^{+\rme^{\mi\theta}\infty} \diff s \, \sqrt{s} \, \text{exp}\le( - \frac{s}{\gym^2} \ri) \mathcal{G}_4 (\sqrt{s}/4\pi)\,,
\eeq
where $\theta = 2\f$. As was shown in \cite{Russo_2012,Aniceto:2014hoa}, this integral has a divergent asymptotic series
\beq \label{eq:asymptotic_Z}
	2 \gym^3 \cdot \mathcal{W}(\gym, 0) \sim 1 - \frac{45 \, \zeta(3)}{256 \pi^4} \, \gym^4 + \frac{525 \, \zeta(5)}{4096 \pi^6} \, \gym^6 + \cO(\gym^8)\,.
\eeq

It is well known that one can interpret the divergence in \eqref{eq:asymptotic_Z} using the notion of \textit{Stokes phenomena}, and obtain a resummed result through the \textit{Borel resummation}. Briefly and in general terms, whenever one encounters a divergent asymptotic series of the type
\beq \label{eq:gevrey-1}
\mathcal{W}(g)\simeq \sum_{k\ge 0} c_k\,g^{k+\beta}\,,
\eeq
with coefficients diverging as $c_k\sim \Gamma(k+\beta)$ for some number $\beta$, 
it is possible to perform a \textit{Borel-transform}
\beq
\mathcal{B}[\mathcal{W}](s) \coloneqq \sum_{k\ge0}\frac{c_k}{\Gamma(k+\beta)}\,s^{k+\beta-1} \,,
\eeq
living in the Borel $s$-plane, which effectively removes the factorial growth from the coefficients and outputs a series with a finite radius of convergence. It is then possible to integrate the resummation/analytic continuation of this series using a Laplace transform along a particular direction $\arg{(s)}=\theta$ to produce a meaningful resummed expression of the original diverging series
\beq
 \mathcal{S}_\theta\mathcal{W}(g)=\int_0^{+\rme^{\mi \theta}\infty}\mathrm{d}s\,\rme^{-s/g}\,\mathcal{B}[\mathcal{W}](s)\,.
\eeq
However, for the Laplace transform above to be well defined, the contour of integration must not encounter any singularities in the Borel plane. When it does - along rays known as \textit{Stokes lines} - one must account for non-perturbative effects. Practically, this means that the Laplace transforms above and below the Stokes line at some $\arg(s) = \theta$ are related by a non-perturbative difference. This difference corresponds exactly to the discontinuity of the original re-summed series along the analogous direction in the $g$-plane
\beq
\mathcal{S}_{\theta_+}\mathcal{W}-\mathcal{S}_{\theta_-}\mathcal{W}=-\mathcal{S}_\theta\,\mathrm{Disc}_\theta\,\mathcal{W}(g)\,,
\eeq
where a Laplace summation of the discontinuity might be needed if the latter is also asymptotic. For in-depth reviews of Stokes phenomena, we refer the reader to \textit{e.g.} \cite{sauzin_lecture_notes, Aniceto_primer}.

As shown in \cite{Aniceto:2014hoa}, the Stokes line associated to the series \eqref{eq:asymptotic_Z} lies along the negative real axis of $s$, with poles in the Borel plane at $- 16 \pi^2 \, n^2$, $n\in \mathbb{N}$,
and the associated discontinuity is a sum of the following form
\beq \label{eq:W_discontinuity}
	2\gym^3\cdot\mathrm{Disc}_\pi \mathcal{W}(\gym, \theta)=2\pi \mi \sum_{n\in\mathbb{N}}\text{exp}\le( \frac{16 \pi^2}{\gym^2} n^2 \ri) \sum_{\ell = 0}^{4n - 1}\frac{(-1)^{4n - \ell - 1}}{\Gamma(4n - \ell)\ell!} \, f_{n,\ell} \, \gym^{-2(4n - 1 -\ell)}\,.
\eeq
We see that the individual terms in this series match with the residues seen in  \eqref{eq:residues}, as expected.\footnote{We saw in sec. \ref{sec:Higgs} that these non-perturbative terms are associated to Seiberg-Witten monopoles. This answers the open question presented in \cite{Aniceto:2014hoa} as to the physical interpretation of the complex saddles seen by their analysis.} We can now formally introduce the coefficients in \eqref{eq:residues}, as the Laurent series coefficients of the Borel transform defined in \eqref{eq:W}, and originally in \cite{Aniceto:2014hoa}
\beq
    \cB[\mathcal{W}](s)|_{s_n} = \frac{1}{(s - s_n)^{4n}} \sum_{\ell = 0}^\infty \frac{f_{n,\ell}}{\ell!}\, (s - s_n)^\ell\,.
\eeq
Their relation to the corresponding coefficients, $\hat{f}_{n,\ell}$ in the $a$--variable, defined in \eqref{eq:series_I_no_exp} is given in appendix \ref{app:Residues}. As an example of their form, the leading-order coefficients, $f_{n,0}$, are
\beq
    f_{n,0} = \mi\cdot n^{1+4n}2^{3+24n}\pi^{8n+1/2}\text{exp}\le[\sum_{k=1}^{n}(12k-8n)\text{log}(k)-\sum_{k=n+1}^{2n}(4k-8n)\text{log}(k)\ri]\,.
\eeq
Higher order coefficients involve corrections in $\gym^{-2}$ and polynomials in the Riemann $\z$-function.

Consider now the two half-contours defined in eq. \eqref{eq:integral_splitting} in terms of the Laplace variable
\beq \label{eq:Z_discontinuity}
\begin{aligned}
    \Z(\gym,\theta) &= \frac{1}{\gym^3\sqrt{\pi}}\le[\int_0^{\rme^{\mi \theta} \infty} \diff s \, \sqrt{s} \, \text{exp}\le( - \frac{s}{\gym^2} \ri) \mathcal{G}_4(\sqrt{s}/4\pi) \ri. 
    \\
    &\phantom{=\;~{\frac{1}{\gym^3}}}-\le. \int_0^{\rme^{-\mi \theta} \infty} \diff s \, \sqrt{s} \, \text{exp}\le( - \frac{s}{\gym^2} \ri) \mathcal{G}_4(\sqrt{s}/4\pi)\ri]
    \\
    &= \frac{1}{\gym^3}(\mathcal{S}_{\theta}\mathcal{W}(\gym,\theta)- \mathcal{S}_{-\theta}\mathcal{W}(\gym,-\theta))
    \\
    & = \frac{1}{\gym^3}\text{Disc}_\pi \mathcal{W}(\gym, \theta)\,.
\end{aligned}
\eeq
Following the discussion of sec. \ref{sec:Higgs}, these equalities hold for the resummed quantities. When $\theta = \pi + \e$ and $\gym \in \mathfrak{D}$, then the resummed Disc$_\pi\mathcal{W}(\gym,\theta)$ can be exchnaged for the formal sum over residues shown in eq. \eqref{eq:W_discontinuity}. Thus, we see that the contour integrals defined in eq. \eqref{eq:integral_splitting} correspond to the discontinuity obtained when crossing the Stokes line at arg$(s) = \pi$. This is equivalent to the statement that the perturbative series stemming from each half of the contour mutually cancel, leaving only the non-perturbative data. Equivalently, the map $a_- \to a_+$ yields a non-trivial monodromy of the function $+\sqrt{s} \to -\sqrt{s}$ in the Borel plane. We therefore find that, while $\mathcal{W}(\gym,\theta)$ is genuinely resurgent, the net contribution to the path integral is
\beq 
    \gym^3 \cdot \Z(\gym) = \text{Disc}_{\phi = \pi}\mathcal{W}(\gym)\,,
\eeq
which is the statement that the path integral is equal to the analytic continuation of the sum over residues, as was deduced in sec. \ref{sec:Higgs}.

We now find a clue as to the physical interpretation of this discontinuity by noting the following: the principal component of Laplace transforms containing only poles can be expressed as a sum of complementary error functions, $\mathrm{erfc}(x)$. 
These are simple resurgent functions, and contain only a single, simple pole in their Borel plane. We can therefore use error functions as a basis for producing resurgent functions that contain only poles in the Borel plane. To see this, note that the function erfc$(x)$ admits the following asymptotic series for $\mathfrak{Re}[x]\to\infty$ \cite{NIST:DLMF}:
\beq
	\frac{1}{\sqrt{\pi} \, x} - \rme^{x^2}\text{erfc}~(x)\sim\frac{1}{x\sqrt{\pi}}\sum^\infty_{n=0}(-1)^{n + 1}\frac{(2n+1)!!}{(2x^2)^{n + 1}}=: \, \Phi(x)\,.
\eeq
Using the natural expansion variable as $z=x^2$ we can write
\beq
\Phi(z) = \sum_{n=0}^\infty{a_n} \, z^{-n-3/2}\,,\quad a_n=\,\frac{(2n+1)!!}{\sqrt{\pi}(-2)^{n+1}}\,.
\eeq
This is an asymptotic expansion of the type \eqref{eq:gevrey-1} where $\beta = 1/2$. The factorial divergence is explicit in the coefficients, with a simple enough series such that the Borel transform resums to
\beq
    \cB[\Phi](s) \equiv \sum_{n = 0}^{\infty} \frac{(-1)^{n+1}}{\sqrt{\pi}} \frac{(2n + 1)!!}{2^{n + 1} \G(n + 3/2)}s^{n + 1/2} = -\frac{1}{\pi}\frac{\sqrt{s}}{1 + s}\,,
\eeq
which has poles at $s = -1 $. Higher-order poles in the Borel plane can be obtained by taking derivatives, which are then mapped to polynomials in the physical plane via a Laplace transform. In the language of resurgence this is a map from the convolutive (Borel) model to the multiplicative (physical) model (see \cite{Aniceto_primer} for a detailed discussion). A sum of error functions with appropriately chosen coefficients will produce the correct singularity structure in the Borel plane of $\Z$ -- a pole of the same order and with the same residue. Thus, the non-perturbative information contained in $\Z$ can equivalently be seen as the non-perturbative data of a set of error functions.

More explicitly,
\beq    
    \frac{\gym}{4\pi^{3/2} n} - \exp\left(\frac{16\pi^2n^2}{\gym^2}\right)\text{erfc}~\left(\frac{4\pi n}{\gym}\right)\longrightarrow \frac{1}{\pi}\int \exp\left(-\frac{16 \pi^2 n^2}{\gym^2} s\right)\frac{\sqrt{s}}{1 + s} \, \diff s\,.
\eeq
Changing variables $s \to s/16\pi^2 n^2$, we obtain a pole at the right position in the Borel plane. Taking $4n - 1$ derivatives of the Borel transform then produces a pole of the correct degree.
 As we mentioned, derivatives in the convolutive model map to a monomial of the same power in the multiplicative model:
\beq
    \frac{d^k}{ds^k}\mathcal{B}[\Phi](s)\longrightarrow \gym^{-2k}\Phi\,,
\eeq
and so we can produce a given pole of degree $k$ in the integrand of $\Z$ by multiplying an error function by $\gym^{2k}$. Finally, for a given pole of leading order $4n$ in the integrand of $\Z$, there are subleading poles up to simple ones. These have coefficients $f_{n,l}$, where $f_{n,0}$ multiplies the pole of order $4n$, $f_{n,1}$ the pole of order $4n-1$ etc. So, in general, a finite polynomial of order $4n-1$ must multiply a given error function to produce the precise singularity structure of the Coulomb branch integral
\beq
    -\mi\pi\le[\frac{\gym}{4 \pi^{3/2} n} -  \exp\le(\frac{16\pi^2n^2}{\gym^2}\ri)\text{erfc}\le(\frac{4\pi n}{\gym}\ri) \ri]\,\sum_{\ell=0}^{4n-1}\frac{(-1)^{4n - 1 -\ell} f_{n,\ell}}{\G(4n - \ell)\ell!}~\gym^{-2(4n - 1 - \ell)}\,.
\eeq
Adding this to the equivalent expression obtained from the negative contour and summing over all integers $n \in \mathbb{N}$ would reproduce the full discontinuity of $\Z$ seen in \eqref{eq:W_discontinuity}, requiring an infinite number of error functions.

This behavior is \textit{generically true for any localization result that contains only poles}. Furthermore, the appearance of error functions hints at a deeper structure underlying the residue sum. As shown by Witten in \cite{Witten:1992xu}, error functions arise via non-Abelian localization as exact formulae for the weak coupling expansion of 2d Yang-Mills gauge theory. What we find is similar: the non-perturbative data of the infinite sum of error functions is a valid representation of $\Z$ when $\gym$ is suitably small in magnitude. We take this as an indication that the Higgs-localized form of $\Z$ can be interpreted in terms of an appropriate two-dimensional theory.

\section{On chiral algebras and unstable configurations}
\label{sec:pole_physics}

In this section, we describe a well-motivated conjecture for the physical origin of the non-perturbative effects we have been studying. In \ref{subsec:clues_and_connections}, we recapitulate the main points regarding our results for the partition function from previous sections. We briefly review relevant results in the literature, specifically those linking theories in four and two dimensions. In sec. \ref{sub:the_2d_model}, we present our 2d model and compare its partition function to the one in 4d. 

\subsection{\label{subsec:clues_and_connections}Motivation and connections to previous work}
We have seen that the Borel-plane-like localized path integral for
the $\mathbb{S}^{4}$ partition function of a 4d $\mathcal{N}=2$
SU$\left(2\right)$ gauge theory with matter yields a finite expression
in two mutually exclusive regions for the coupling in the complex plane. Moreover, these expressions
are simultaneously valid, i.e. true for a single theory, precisely when
the theory in question possesses quantum superconformal invariance.
The first expression is given by integration along the conventional
contour for the bosonic field of the theory and is well defined, for instance for
$g_{\text{YM}}$ real, as long as the theory is asymptotically free (or
finite). Alternatively, for $g_{\text{YM}}$ with small modulus and
argument in a region around $\pi/2$ (or $-\pi/2$), the integral
is well defined along a contour for the bosonic scalars close to the
imaginary axis for some theories. Integration along the latter contour
can also be recast as an infinite sum over the residues of the integrand,
which lie along the imaginary axis. We have argued that when both
expressions are valid the second expression can be thought of as an
analytic continuation of the first. 

The situation described above, a Borel-plane integral with only poles,
is reminiscent of the exact expressions for the weak coupling expansion
of pure 2d Yang-Mills theory, derived using localization in \cite{Witten:1992xu, Griguolo:2024ecw}. In this section, we would like to argue for a direct connection
between these two classes of theories. While pure 2d Yang-Mills is
not supersymmetric, it can be presented as an A-twisted $\left(2,2\right)$
2d gauge theory, and some of its observables, including the partition function on a Riemann surface of arbitrary genus, can be computed via supersymmetric localization  \cite{Witten:1992xu, Griguolo:2024ecw,Leeb-Lundberg:2023jsj}. 
In fact, the partition function of pure 2d Yang-Mills theory
can be computed exactly by a variety of methods, including also a lattice setup \cite{Migdal:1975zg,Witten:1991we}, and a specific type of gauge fixing \cite{Blau:1993hj}. In all but the lattice treatment, the partition function is saturated by unstable instanton configurations associated with connections satisfying the 2d Yang-Mills equations. Such configurations are generically not topologically protected. See \cite{ Witten:1992xu, Atiyah:1982fa, Griguolo:2024ecw} for a discussion of these configurations. 

There are also a variety of known relationships between supersymmetric gauge theories in 4d and theories in 2d. Since the proposed setup is only applicable for superconformal theories, we will concentrate on two such relationships which apply to this class. 

\paragraph{$\mathcal{N}=4$ SYM and complex 2d YM.} The first relationship is the reduction of certain computations in $\mathcal{N}=4$ SYM, for instance the expectation value of $1/8$ BPS Wilson loops, to computations in a version of 2d Yang-Mills, first conjectured in \cite{Drukker_2008b,Drukker_2007,Drukker_2008a}. The conjecture implies an unusual relationship between the 4d gauge coupling $g_{\text{YM}}$ and the gauge coupling $g_{\text{2d}}$ of the putative 2d theory, of the form 
\beq
g_{\text{2d}}^{2}=-\frac{g_{\text{YM}}^{2}}{4\pi}\,.
\eeq
This means that for real $g_{\text{YM}}$ the 2d theory has an imaginary coupling. A similar relationship is implied by our results in sec. \ref{sub:the_2d_model}, but the interpretation is different: the 2d theory is an ordinary gauge theory and the observables computed in the 2d theory are the analytic continuation of those of the 4d theory.  $\mathcal{N}=4$ SYM is the simplest example of a 4d $\mathcal{N}=2$ superconformal theory, and it would be nice
to reconcile these two points of view. We will not do so fully,
but only point out that, in the presence of $\mathcal{N}=4$ supersymmetry, the integrand coming from localization for the partition function no longer has any poles anywhere in the complex plane. In the context of our analysis this indeed implies that no instanton contributions are present in the 2d theory.

A derivation of the relationship, using supersymmetric localization and incorporating a plausible conjecture regarding one-loop contributions, was put forward in \cite{Pestun_2012}. The analysis in \cite{Pestun_2012} claims that the 2d model which should be used to compute the 4d Wilson loops is a version of 2d Yang-Mills theory, the Hitchin/Higgs-Yang-Mills theory, further restricted to the zero instanton sector. The restriction happens dynamically, through the existence of certain fermion zero modes. While the exact setup used in \cite{Pestun_2012} is not available for other superconformal $\mathcal{N}=2$ theories, it is reasonable to conjecture that a similar localization computation can be used to derive an analogous relationship in those cases, perhaps incorporating a more sophisticated 2d model. In sec. \ref{sub:the_2d_model}, we will provide preliminary evidence that this is indeed so. 

\paragraph{The chiral algebra construction.} In \cite{Beem_2015}, the authors derive a different, more general relationship between any superconformal 4d $\mathcal{N}=2$ theory and chiral algebras, which we very briefly review below for the simple case of Lagrangian theories. Specifically, the authors identified a distinguished subset of the local operators of any superconformal 4d $\mathcal{N}=2$ theory which form a chiral algebra. While not fully-fledged QFTs themselves, these chiral algebras may form parts of a 2d CFT.  The construction works as follows. When restricted to lie in a specific plane within $\mathbb{R}^{4}$, 4d superconfomal symmetry guarantees that a subset of the plane translations of these operators are exact under a specific supercharge. The supercharge in question is a combination of a Poincaré and a conformal supercharge, and is not conserved in non-conformal theories. The authors moreover showed that the OPE of these particular local operators coincides with that of specific chiral algebras in the plane, and computed some of the invariants of these algebras. As it turns out, the central charge resulting from this construction is always negative and the chiral algebra in question is non-unitary. 

According to \cite{Beem_2015}, a free hypermultiplet in a 4d superconformal $\mathcal{N}=2$ theory gives rise to the chiral algebra of a symplectic boson. The relevant operators are specific specetime-dependent linear combinations of the bosons in the hypermuliplet. Similarly, a free vector multiplet gives rise to a $\left(b,c\right)$ ghost system comprising combinations of gauginos. The existence of gauge symmetry also serves to reduce the content of the algebra to the gauge invariant operators. However, the chiral algebra construction is independent of the gauge coupling and, in fact, all marginal parameters. We refer the reader to \cite{Beem_2015} for more details. 

In \cite{Pan:2017zie}, the authors argued that the relationship of a superconformal 4d theory to the chiral algebra, as described above, could be used to give an alternative expression for expectation values of operators in this algebra, computed on the four sphere using supersymmetric localization. Specifically, the computation utilises an alternative localizing supercharge compatible with the chiral algebra construction. Computations of the expectation values of operators in the chiral algebra are reduced to computations in an effective 2d theory. Some of the properties of this 2d theory were derived in \cite{Pan:2017zie}, closely following the derivation in \cite{Pestun_2012}, while others seemed puzzling. For instance, the expectation values certainly do depend on the marginal Yang-Mills coupling. In sec. \ref{sub:the_2d_model}, we will argue that, with appropriate modifications, the partition function for the theory, i.e. the expectation value of the unit operator, can indeed be computed in this way. We will comment only briefly on the rest of the chiral algebra. 

\subsection{\label{sub:the_2d_model}The 2d model}

We now make the minimal hypothesis for a 2d theory which computes the partition function of a superconformal $\mathcal{N}=2$ gauge theory on the four sphere in the regime where the sum over residues is valid. The theory is simply 2d Yang-Mills, on the round two sphere, minimally
coupled to the chiral bosons and $\left(b,c\right)$ ghost systems implied by the chiral algebra construction. The existence of the chiral algebra elements was demonstrated in \cite{Pan:2017zie}, while the complete setup, including the incorporation of the coupling, is novel. 

The matter in our theory is conformal, hence there is a canonical way of putting it on the two sphere, which is insensitive to the radius of the sphere. Similarly, 2d Yang-Mills is invariant under area-preserving differomorphisms
and putting it on the two sphere only involves fixing the ratio of the Yang-Mills coupling $g_{\text{2d}}$ to the radius of the sphere. For simplicity we will choose this radius to be $1$ and continue to denote the ratio by $g_{\text{2d}}$. We will see that in order for the computation of the the partition function in the 2d theory to match the 4d result, it must be analytically continued to imaginary values of the coupling, such that $g_{\text{2d}}=-\mi \, g_{\text{YM}}/2$. 

\paragraph{Promotion to a $\left(0,2\right)$ theory.}

In order to compute the partition function of our 2d theory, we will follow a procedure similar to the one described in \cite{Witten:1992xu} for pure 2d Yang-Mills. That is, we will promote the theory to a twisted supersymmetric gauge theory, and argue that the partition function remains the same. For pure 2d Yang-Mills, an appropriate choice of theory is an A-twisted $\left(2,2\right)$ gauge theory with no additional
matter. One could certainly add $\left(2,2\right)$ matter to this setup and repeat the computation, as was done in  \cite{Leeb-Lundberg:2023jsj}. However, in the case at hand neither the matter content, nor indeed the result of the localization computation, are compatible with $\left(2,2\right)$ supersymmetry. Moreover, the analysis of the chiral algebra in \cite{Beem_2015} and its four sphere incarnation in \cite{Pan:2017zie} strongly imply that any associated 2d theory should have chiral supersymmetry . 

We therefore consider a 2d $\left(0,2\right)$ gauge theory. We will consider \textit{only} the A/2-twisted version of the theory. We refer the reader to \cite{Closset_2016} for details regarding the multiplet structure and twisting in this class of theories. Since the theory has chiral supersymmetry, the twist does not result in a topological field theory. Instead, the partition function of the twisted theory depends on the metric via its conformal class.\footnote{The conformal anomaly will not play a role in our analysis.} Moreover, expectation values of local operators in the twisted
cohomology have a dependence on their insertion points that is meromorphic in the complex structure associated to that class. This is the same as the metric dependence of a 2d CFT with chiral operator insertions. We will consider only the partition function of the theory on the two-sphere, in which case both types of dependence are trivial.

\begin{table}
\begin{centering}
\begin{tabular}{|c|c|c|c|}
\hline 
\textbf{$\left(0,2\right)$ multiplet type} & \textbf{field content} & \textbf{twisted spins} & \textbf{gauge representation}\tabularnewline
\hline 
\hline 
vector & $a_{1},a_{\bar{1}},\psi_{1},\psi_{0},D$ & $\left(1,-1,1,0,0\right)$ & adjoint\tabularnewline
\hline 
chiral & $\phi,\mathcal{C}$ & $\left(0,-1\right)$ & adjoint\tabularnewline
\hline 
anti-chiral & $\bar{\phi},\mathcal{B}$ & $\left(1,1\right)$ & adjoint\tabularnewline
\hline 
Fermi & $\lambda_{1},\mathcal{G}_{0}$ & $\left(1,0\right)$ & adjoint\tabularnewline
\hline 
Fermi & $\lambda_{0},\mathcal{G}_{\bar{1}}$ & $\left(0,-1\right)$ & adjoint\tabularnewline
\hline 
anti-Fermi & $\bar{\lambda}_{1},\mathcal{\bar{G}}_{1}$ & $\left(1,1\right)$ & adjoint\tabularnewline
\hline 
anti-Fermi & $\bar{\lambda}_{0},\bar{\mathcal{G}}_{0}$ & $\left(0,0\right)$ & adjoint\tabularnewline
\hline 
chiral & $q_{A},\mathcal{C}_{A}$ & $\left(1/2,-1/2\right)$ & $\mathfrak{R}\oplus\bar{\mathfrak{R}}$\tabularnewline
\hline 
anti-chiral & $\bar{q}_{A},\bar{\mathcal{B}}_{A}$ & $\left(1/2,1/2\right)$ & $\mathfrak{R}\oplus\bar{\mathfrak{R}}$\tabularnewline
\hline 
chiral & $u,\mathcal{C}_{u}$ & $\left(1/2,-1/2\right)$ & adjoint\tabularnewline
\hline 
anti-chiral & $\bar{u},\mathcal{B}_{\bar{u}}$ & $\left(1/2,1/2\right)$ & adjoint\tabularnewline
\hline 
Fermi & $v,\mathcal{G}_{v}$ & $\left(0,-1\right)$ & adjoint\tabularnewline
\hline 
anti-Fermi & $\bar{v},\bar{\mathcal{G}}_{v}$ & $\left(1,1\right)$ & adjoint\tabularnewline
\hline 
\end{tabular}
\par\end{centering}
\caption{Field content of the twisted $\left(0,2\right)$ gauge theory.}\label{tab:field_content}
\end{table}

Our proposed field content is given in Table \ref{tab:field_content}. Note that we have assigned slightly unorthodox quantum numbers to the anti-chiral and anti-Fermi multiplets, which would usually have opposite quantum numbers from their conjugate multiplets. This is necessary in order to reproduce the four dimensional result. It is one of several indications that the reality conditions in this setup are delicate. We believe that the field content, including these quantum numbers, and the action discussed below can be derived from the 4d theory. Indeed, some of these were already identified in \cite{Pan:2017zie}. 

\paragraph{Anomalies.}

Unlike models with $\left(2,2\right)$ supersymmetry, 2d $\left(0,2\right)$ gauge theories can have gauge anomalies, of both abelian and non-abelian type. They may also have mixed gauge-R-symmetry anomalies. Since our models descend from UV complete 4d gauge theories, the gauge group is semi-simple and only the non-abelian gauge anomaly can be present. For simplicity, we present the anomaly in the case when the gauge
group is simple. In the notation of \cite{Closset_2016}, with $i$ indexing chiral multiplets and $I$ Fermi multiplets, the anomaly is proportional to 
\beq
\sum_{i}T_{\mathfrak{R}_{i}}-\sum_{I}T_{\mathfrak{R}_{I}}-T_{\text{adj}}\,,
\eeq
where $T$ denotes the Dynkin index of the representation $\mathfrak{R}$. For representation
matrices $t_{\mathfrak{R}}^{a}$, the Dynkin index can be defined
by the relation 
\beq
\text{Tr}_{\mathfrak{R}}\left(t_{\mathfrak{R}}^{a}t_{\mathfrak{R}}^{a}\right)=T_{\mathfrak{R}}\delta^{ab}\,,
\eeq
and specifically $T_{\text{adj}}=h^{\vee}$, the dual Coxeter number.
For SU$\left(N\right)$ we have
\beq
T_{\text{fund}}=T_{\text{anti-fund}}=\frac{1}{2}\,,\quad T_{\text{adj}}=N\,.
\eeq

For the matter content described above, cancellation of the non-abelian
gauge anomaly therefore requires
\beq
T_{\mathfrak{R}}+T_{\mathfrak{\bar{R}}}-2T_{\text{adj}}=0\,.
\eeq
Note that this is exactly the same condition as the one which guarantees
the vanishing of the beta function of a 4d $\mathcal{N}=2$ gauge
theory, and therefore 4d superconformal invariance.\footnote{\textit{c.f.} equation (3.1.6) in \cite{Tachikawa:2013kta}.}
Hence, convergence
of the sum over poles, as discussed in \ref{sec:Higgs},  and anomaly cancellation of the 2d gauge theory, both require the 4d $\mathcal{N}=2$ theory
to be superconformal. We take this as an encouraging sign that the
two models are related.

\paragraph{The partition function.}

In order to make contact with the results in previous sections, we
now mostly restrict ourselves to gauge group SU$\left(2\right)$ and $N_{f}$
chiral multiplets, those with bottom components $q_A$, in the fundamental and anti-fundamental
representation. The restriction from anomaly cancellation,
or conformal invariance, becomes
\beq
N_{f}=2N_{c}=4\,.
\eeq

We will denote the action of the twisted supersymmetry by $\delta$. Our supersymmetric action will consist of some classical $\delta$-closed terms, together with arbitrary $\delta$-exact terms to be discussed later on. An appropriate classical action for the fields comprising the chiral algebra, in terms of the lowest components in the superfield, on an arbitrary Riemann surface is 
\beq
    \int \left(\Omega^{AB}\,q_A \bar{\partial}q_B  +\text{tr}\left(\lambda_1\bar{\partial}\lambda_0\right)\right)\,
\eeq
where $\Omega^{AB}$  is the symplectic form which yields a gauge invariant bilinear for the hypermultiplet and $\bar{\partial}$ is the twisted (gauge-covariant) Dolbeault operator, denoted in \cite{Closset_2016} by $D_{\bar{1}}$. The classical action for the gauge sector we take to be
\beq \label{classical_Yang_Mills_term}
    \int \text{tr}\left(i\phi f_{1\bar{1}}+\frac{i}{2}\mathcal{C}\psi_1-\epsilon\phi^2\right)\,,
\eeq
with $\epsilon$ an arbitrary positive constant which we will eventually choose so that the result coincides with our results in 4d. This is the same as the classical action considered in \cite{Griguolo:2024ecw,Leeb-Lundberg:2023jsj,Leeb-Lundberg:2025baz}. As shown there, integrating out the multiplet whose bottom component is $\phi$, we recover our original non-supersymmetric action consisting of chiral matter coupled to 2d Yang-Mills. 

We now employ supersymmetric localization by adding $\delta$-exact actions for the rest of the multiplets. We will not attempt to track the overall normalisation of the result, which in any case differs from the standard one, only the dependence on the Yang-Mills coupling constant. 

The relevant localization calculation for the vector multiplet was performed in \cite{Closset_2015}. According to
these authors, the vector multiplet fields localize to a Coulomb branch configuration where the Hodge dual of the field strength, $f_{1\bar{1}}$ in complex coordinates, satisfies
\beq
\frac{1}{\pi}\int\sqrt{g}f_{1\bar{1}}=\mathfrak{m}\,,
\eeq
with $\mathfrak{m}$ belonging to the GNO lattice for the gauge group. 

The rest of the matter content does not belong to complete $\left(2,2\right)$ multiplets and lies outside the boundaries of the analysis performed in \cite{Closset_2015} and \cite{Closset_2016}. Specifically, the localizing terms considered in these references are not available. However, the fields still admit alternative localizing terms. Given a pair of conjugate multiplets, a chiral and an anti-chiral or a fermi and anti-fermi, with components $X,Y$ and $\bar{X},\Bar{Y}$ respectively, our localizing terms will be of the form

\beq
\delta\int\sqrt{g}\,\bar{X}\,Y=\int\sqrt{g}\left(\bar{Y}Y\pm 2i \bar{X}\bar{\partial}X\right)\,.
\eeq
The sign in the kinetic term corresponds to chiral vs Fermi multiplets, respectively. For instance, for the chiral multiplets $q_A$, this term is 
\beq
\delta\int\sqrt{g}\,\bar{q}^A\,\mathcal{C}_A=\int\sqrt{g}\,\left(\bar{\mathcal{B}}^{A}\mathcal{C}_{A}+2i\bar{q}^A \bar{\partial}q_A \right)\,.
\eeq
Note that the holomorphic part of the covariant derivative does not appear, and that there is
no kinetic term for the top components. These differences mean that the localization results in this context, specifically the one-loop effective action coming from integrating out these fields, are somewhat different from those in the models presented in e.g. \cite{Closset_2016}.  

Among the constant modes of the field $\phi$, those which are valued in the same Cartan subalgebra as $\mathfrak{m}$ are true zero modes of the localizing action. Integrating them out results, as before, in a classical Yang-Mills action, now of the localized form  
\beq
    S_\text{classical}=\frac{16\pi^2}{g_\text{2d}^2}\mathfrak{m}^2\,,
\eeq
where we have made a specific choice for the coupling constant. The rest of the fields will be treated in the one-loop approximation. Since the fields are charged, the effective spin of a component of weight $\rho$ under the gauge group is shifted by $\rho\left(\mathfrak{m}\right)/2$ from its original values. For instance, each of the bosons $q_A$, which are in the fundamental/anti-fundamental representation, have twisted
spins $\mathfrak{m}/2+1/2$ and $-\mathfrak{m}/2+1/2$, while the adjoint fermions $\lambda_0,\bar{\lambda}_0$
have components with twisted spins $\pm\mathfrak{m}$ and $0$.

\paragraph{Eigenmodes and eigenvalues}

The one-loop determinants which follow from these localizing actions
can be evaluated using spin-weighted (monopole) spherical harmonics
and the relationship between the spin-weighted partial differential
operator $\eth$ and the Dolbeault operator. Monopole spherical harmonics
$_{s}{Y}_{lm}$ are sections of complex line bundles over $\mathbb{C}P^{1}$ which are non-vanishing
for all $\left|s\right|\leq l$ and $-l\leq m\leq l$ and satisfy
\beq
\int\sqrt{g}{}_{s}{Y}_{lm}\,{}_{-s}{Y}_{l'-m'}=e^{i\alpha}\delta_{ll'}\delta_{mm'}\,,
\eeq
with $\alpha$ a convention-dependent phase. We will need the following result for the action of the $\bar{\partial}$ operator
\beq
\bar{\partial}{}_{s}{Y}_{lm}=-\sqrt{\left(l+s\right)\left(l-s+1\right)}{}_{s-1}{Y}_{lm}\,.
\eeq
After choosing an isomorphism between the spaces mapped by $\bar{\partial}$, which is related to choosing a contour of integration for the fields, the determinant of this operator can be evaluated using these generalized eigenfunctions and eigenvalues.\footnote{As previously indicated, the contour of integration in our model is subtle. We will comment more on this point in the discussion. See also the discussion of reality conditions in \cite{Pan:2017zie}. }  A detailed discussion of the relevant regularised determinants can be found in \cite{bar2003dirac}, however we will present a self-contained treatment which allows us to relate them to the 4d result. 

Evaluating the quadratic integrals for the localizing actions yields a product of the generalized eigenvalues above. However, this product is \textit{infinite} due to zero modes for some fields, e.g. the fields $q_A$. Worse, there are ambiguous $0/0$ expressions associated with simultaneous zero modes of bosons and Fermions. 

In \cite{Closset_2016}, the existence of a continuous modulus for $\phi$, which coupled to these fields,  yields an effective mass for charged modes and avoids this catastrophe. Instead of ambiguous expressions, one has poles at specific positions in the moduli space for $\phi$. However, no such coupling is present in our setup, nor can it be introduced for e.g. the fields $q_A$. Comparing with the setup in \cite{Closset_2016}, we are evaluating the one loop determinants at the exact position of their poles. 

In the context of the more standard twisted theories of \cite{Closset_2016,Closset_2015}, integration over the continuous modulus could be shown to require taking the \textit{residue} of the effective action, comprising both the one-loop determinants and the classical terms, at the position of poles. Indeed, the partition function is evaluated in these setups using the JK residue prescription \cite{Benini:2013nda,Benini:2013xpa,jeffrey1994localizationnonabeliangroupactions}. We do not have such a continuous modulus. Nevertheless, we now argue that taking the residue is still the correct procedure by regularizing the situation in an alternative way.  

The sections ${}_{s}{Y}_{lm}$  can be sensibly continued to non-integer values of $s$ and $m$, for instance by using their expressions as hypergeometric functions. For consistency, this continuation of the eigenvalues should be done all at once in both the classical and one-loop terms. For an appropriate continuation, defined in this case by shifting $s,l$, and $m$ by the same negative amount $\epsilon$ , we get the following:
\begin{enumerate}
\item Eigenfunctions for $l<|s|$  no longer vanish, and have purely imaginary generalized eigenvalues which do not vanish in the limit $\epsilon\rightarrow 0$.
\item  Eigenfunctions for $l=|s|$  and $s<0$, which were previously zero modes, now have purely imaginary eigenvalues which vanish in the limit $\epsilon\rightarrow 0$.
\item All other eigenfunctions have non-zero eigenvalues which are shifted slightly away from their original positions, and whose original values are recovered in the limit $\epsilon\rightarrow 0$.
\end{enumerate}
We will regularize our expression for the partition function using $\epsilon$, which therefore means including the additional eigenfunctions and then taking the limit $\epsilon\rightarrow 0$ carefully at the end. 

The regularization we propose requires evaluating the Gaussian integrals for would-be zero modes which, after regularization, have purely imaginary eigenvalues proportional to $\epsilon$. As shown in \cite{Griguolo:2024ecw,Leeb-Lundberg:2023jsj,Leeb-Lundberg:2025baz}, the correct procedure in such a situation is to replace the contribution of the zero modes with a \textit{distribution} depending on $\mathfrak{m}$ and $\epsilon$ with singular support at $\epsilon=0$. The appropriate distribution, including the normalization, depends on the details of the matter content. We will adopt the conclusions of \cite{Leeb-Lundberg:2025baz}, which addresses twisted $\left(2,2\right)$ theories with chiral multiplets, by taking the distribution to be an appropriate number of derivatives of the Dirac delta function. The number of derivatives is fixed by the the difference between the number of zero modes for bosonic fields and the same number for Fermionic fields, denoted as $n\left(\mathfrak{m}\right)$: this is the overall divergence of the one-loop determinant at a fixed $\mathfrak{m}$. We further take the value of the result to be the infinitesimal integral around $\epsilon=0$ implied by the delta function, rather than the pointwise limit $\epsilon\rightarrow 0$. We refer the reader to \cite{Griguolo:2024ecw}  for a derivation of the distribution and its relationship to the effective number of zero modes, and to \cite{Leeb-Lundberg:2025baz}  for a closely related result in the context of twisted $\left(2,2\right)$ gauge theories with chiral multiplets. 

Denoting the original expression for the effective action as $Z\left(\mathfrak{m}\right)$, our regularized expression is therefore

\beq\label{eq:zero_mode_residue}
    \lim_{\epsilon\rightarrow 0}\,\frac{1}{\left(n\left(\mathfrak{m}\right)-1\right)!} \frac{d^{n\left(\mathfrak{m}\right)-1}}{d\epsilon^{n\left(\mathfrak{m}\right)-1}}\left(\epsilon^{n\left(\mathfrak{m}\right)}Z\left(\mathfrak{m}+\epsilon\right)\right)=\text{Res}_{\epsilon\rightarrow 0}Z\left(\mathfrak{m}+\epsilon\right)\,.
\eeq
Although the numerical factors in this expression appear naturally in the analysis of \cite{Griguolo:2024ecw,Leeb-Lundberg:2025baz}, we have not attempted to fix the overall normalisation. 

\paragraph{One-loop determinants}
Our next task is to compute the 1-loop determinants for the $\d$--exact actions introduced at the start of this section. To do so, we expand the fields of the matter multiplets in terms of our analytically continued monopole spherical harmonics. This yields the following generic form of one-loop determinants as a function of the effective twisted spin, $s = s_0 + \rho(\fm)/2$
\beq \label{eq:unreg_1loop}
Z_s = \prod_{\rho\in\mathfrak{R}_i}\prod_{l=\left|s\right|}^{\infty}\left(\left(l+s\right)\left(l-s+1\right)\right)^{l+1/2}\,.
\eeq
The \textit{bare spin}, $s_0$, is defined as the spin in the trivial background, when $\fm = 0$. The computation of the 1-loop determinant for the symplectic boson was performed in \cite{Pan:2017zie}. There, the calculation was deduced directly from the four-dimensional theory with a free hypermultiplet, by localizing the theory with respect to the same supercharge used to identify the relevant chiral algebra in \cite{Beem_2015}. Being the free hypermultiplet, this calculation corresponded to a two-dimensional theory in a trivial flux $\fm = 0$ background, so $s = s_0 = 1/2$, for the corresponding 2d chiral boson. It was also shown in appendix B of \cite{Pan:2017zie}, that the 4d determinant for the vector multiplet and gauged hypermultiplet can be produced from the kinetic actions of a symplectic boson, and a small $(b,c)$ ghost system in the presence of a carefully selected background for the four-dimensional gauge field. In this work, we would like to reproduce these results by obtaining the residues of the full 4d expression, \eqref{eq:Z_SU(2)}, discussed in the previous sections, from a purely 2-dimensional perspective, following methodologies outlined in \cite{Benini_2014_2D, Closset_2015, Closset_2016}.

Our analysis is entirely based on how we have chosen to address the zero modes of \eqref{eq:unreg_1loop}. The 1-loop determinants are computed in monopole backgrounds $\fm$, with effective spin $s = s_0 + \rho(\fm)/2$. Lifting the zero modes by deforming away from the BPS locus by a small real parameter $\fm + \epsilon \equiv \fa$, means that we will treat the effective spin $s = s_0 + \r(\fa)/2$ as a continuation about the bare spin, $s_0$. As explained above, this includes relaxing the restriction that $l \geq |s|$ and we will therefore take $l = s_0 + n,$ $n \in \mathbb{Z}_{\geq 0}$. Strictly speaking, the following determinants should be computed without the zero modes, using the prescription described above to extract the residues. For brevity of notation we will compute the determinants including the zero modes, and take the final result to be multiplied by $\e^{n(\fm)}$, as in \eqref{eq:zero_mode_residue}.

\paragraph{Fundamental chiral multiplets: $\mathbf{s_0 = 1/2}$.} We start by expressing \eqref{eq:unreg_1loop} in the following way
\beq
    Z^{\phi_{1/2}}=\prod_{\rho}\prod_{l=1/2}^\infty((l+1/2)^2 - (s-1/2)^2)^{l+1/2}\,.
\eeq
Upon inserting $s = s_0 + \r(\fa)/2$, where $\fa = \fm + \e$, the 1-loop determinant becomes
\beq \label{eq:ch_1loop}
    Z^{\phi_{1/2}}=\prod_{l = 1/2}^\infty((l+1/2)^2 - \fa^2/4)^{2l + 1}\,.
\eeq
This product is divergent, and so we will employ zeta-function regularization to obtain a finite result. By introducing the multiple Hurwitz zeta function for parameters $a \in \mathbb{C}$, $\underline{w} = (w_1,w_2,\hdots, w_r)>0$, and $\mathbf{n}=(n_1,n_2,\hdots, n_r) \geq 0,\, n_i\in\mathbb{Z}$
\beq
    \zeta_r(s,a,\underline{w})\coloneqq \sum_{\mathbf{n} \geq 0}(\mathbf{n}\cdot \underline{w} + a)^{-s}\,,
\eeq
we can define the multiple gamma functions, $\G_r(a,\underline{w})$ through the following zeta function regularization \cite{multiple_sine_functions}
\beq
    \G_r(a,\underline{w})\coloneqq \text{exp}\le( \frac{\partial}{\p s}\zeta_r (s,a,\underline{w})\bigg|_{s=0} \ri)\,.
\eeq
To compute the 1-loop determinants for the matter fields, we will thus require the zeta function regularization of the following products
\beq
\begin{aligned}
    \prod_{n = 0}^\infty(n + a) &= \phantom{,}\text{exp}\le( \le.\frac{\p}{\p s} \zeta_1(s,a,1)\ri|_{s=0}\ri) \phantom{1}\coloneqq \G_{1}(a,1)^{-1}\,,
    \\
    \prod_{n = 0}^\infty (n + a)^{n+1} &= \text{exp}\le( \le.\frac{\p}{\p s} \zeta_2(s,a,1,1)\ri|_{s=0}\ri) \coloneqq \G_2(a,1,1)^{-1}\,.
\end{aligned}
\eeq
Therefore, products with degeneracy $bn + c$ are regularized to the following
\beq \label{eq:general_zeta_reg}
    \prod_{n=0}^\infty(n+a)^{bn + c} = \G_2(a,1,1)^{-b}\,\G_1(a,1)^{b - c} = \le(2\pi\ri)^{(c-b)/2}G(1 + a)^b \,\G(a)^{-c}\,, 
\eeq
where we have used the fact that $\G_1(z,1)=(2\pi)^{-1/2}\G(a)$ and $G(1+a)$ is the Barnes--G function, introduced in sec. \ref{sec:Higgs} \cite{spreafico_asymptotics}
\beq
    G(1+a) = \G(a) \, \G_2(a,1,1)^{-1}\,.
\eeq
Using this, the determinant \eqref{eq:ch_1loop} regularizes to
\beq
    \prod_{n = 0}^\infty((n+1)^2 - \fa^2/4)^{2n + 2} = \le(\frac{G(2 + \fa/2)G(2-\fa/2)}{\G(1+\fa/2)\G(1-\fa/2)}\ri)^2 = (G(1+\fa/2)G(1-\fa/2))^2\,.
\eeq
The factor is then raised to the power of $N_f$, the number of chiral multiplet pairs, which we may recognize as the same contribution arising from the four-dimensional hypermultiplets.

\paragraph{Adjoint Fermi multiplets: $\mathbf{s_0 = 0}$.} For the quadratic term of the lowest component of the Fermi multiplet $\Lambda$, we have twisted spin $s_0 = 0$. In this case, the determinant is equivalent to 
\beq \label{eq:unreg_1loop_Fermi}
    Z^{\Lambda_0} = \prod_{l = 0}^{\infty}(l+1)^{2l+1} \prod_{l = 0}^\infty(l^2 - \fa^2)^{l+1/2}((l+1)^2 - \fa^2)^{l+1/2}\,.
\eeq
Using the general formula \eqref{eq:general_zeta_reg}, we may evaluate each of these divergent products as follows
\beq \label{eq:Fermi_reg_components}
\begin{aligned}
    \prod_{l = 0}^{\infty}(l+1)^{2l+1} & = \frac{1}{\sqrt{2\pi}}\,;
    \\
    \prod_{l=0}^\infty(l^2-\fa^2)^{l+1/2}&=\frac{1}{(2\pi)^{1/4}}\frac{G(1+\fa)G(1-\fa)}{\sqrt{\G(\fa)G(-\fa)}}\,;
    \\
    \prod_{l=0}^\infty((l+1)^2-\fa^2)^{l+1/2}&=\frac{1}{(2\pi)^{1/4}}\sqrt{\G(1+\fa)\G(1-\fa)}G(1+\fa)G(1-\fa)\,.
\end{aligned}
\eeq
Put together, we find that the 1-loop determinant for an adjoint Fermi-multiplet evaluates to
\beq \label{eq:reg_1loop_Fermi}
    Z^{\Lambda_0}=\frac{\mi}{2\pi} \fa (G(1+\fa)G(1-\fa))^2\,.
\eeq
The factor of $\fa$ appears almost as the Vandermonde determinant. To produce the full Vandermonde, we need to introduce additional Fermi/anti-Fermi and chiral/anti-chiral pairs with twisted spins $0/1$ and $\frac{1}{2}/\frac{1}{2}$, respectively.

\paragraph{Adjoint chiral multiplets: $\mathbf{s_0 = 1/2}$.} This calculation is morally the same as the fundamental chiral multiplets, up to the difference in representations of the gauge group. The product \eqref{eq:unreg_1loop} in this case reads
\beq
    Z^{\Lambda_{1/2}} = \prod_{l=0}(l+1)^{2l+2}\prod_{l=0}((l+1)^2-\fa^2)^{2l+2}= (G(1+\fa)G(1-\fa))^2\,.
\eeq
This therefore provides the necessary cancellation with the $s_0=0$ Fermi multiplet so that we do not introduce any additional zero modes. The overall contribution of the additional Fermi and chiral multiplet pairs is therefore simply a factor of $\mi \fa/2\pi$.

All--in--all, the net 1-loop determinant arising from integrating over quadratic fluctuations of our choices of (0,2) matter multiplets is
\beq \label{eq:4d_1loop_from_2d}
    Z_{\text{matter 1-loop}} = -\frac{\fa^2}{4\pi^2} \frac{(G(1+\fa)(G(1-\fa))^2}{(G(1+\fa/2)G(1-\fa/2))^8}\,.
\eeq
This is precisely the combination of Barnes-G functions seen in \eqref{eq:Z_SU(2)} for $N_f = 4$. Furthermore, despite the BPS loci for the two-dimensional gauge field containing all Yang-Mills connections, $\fm \in \mathbb{Z}$ we see that only even integers will contribute to the residue calculation. This is because the Barnes-G function $G(1+\fa)$ has zeros at $0$ and negative \textit{integers}; if $2\fa_{\epsilon\to0} \in 2\mathbb{Z}+1$ the numerator of \eqref{eq:4d_1loop_from_2d} will have zeros while the denominator will not. These modes therefore evaluate to zero in the residue calculation. We therefore make the identification that $\fm = 2n$ where $n$ is the 4d Chern class of a monopole configuration. Re-introducing the effects of the zero modes, and adding the Vandermonde determinant and classical action, we find
\beq\label{eq:2d_residues}
    Z_{2d} = \underset{\e\to0}{\text{lim}} \frac{1}{(n(\fm)-1)!}\frac{d^{n(\fm)-1}}{d\e^{n(\fm)-1}}\e^{n(\fm)}e^{-\frac{16\pi^2}{g_{2d}^2}(\fm+\e)^2}(\fm+\e)^2 \frac{(G(1+\fm + \e)G(1-\fm - \e))^2}{(G(1+\frac{\fm}{2}+\e) G(1-\frac{\fm}{2}-\e))^8}\,.
\eeq
Noting that $\epsilon = \fa - \fm$, we see that this result exactly matches how the residues were obtained in the original four-dimensional theory, as detailed in appendix \ref{app:Residues}. There, the calculation was performed by hand, by considering the Laurent expansion of the integrand of the 4d Coulomb branch integral. It is interesting to see this result emerge naturally from the 2d localization procedure we have selected. Therefore, we can directly equate eq. \eqref{eq:2d_residues} to the result found in sec. \ref{sec:Higgs}, given in eq. \eqref{eq:residues}. So to summarise, our final result is
\beq
\boxed{
\left.
\begin{aligned}
    Z_{2d} &= \sum_{\fm = 0} \text{Res}_{\e = 0}\le[ \rme^{-\frac{16\pi^2}{g_{2d}^2}(\fm+\e)^2}(\fm+\e)^2 \frac{(G(1+\fm + \e)G(1-\fm - \e))^2}{(G(1+\frac{\fm}{2}+\e) G(1-\frac{\fm}{2}-\e))^8} \ri]
    \\
    &=\frac{1}{2\gym^3}\sum_{n = 0} \text{exp} \le( \frac{16 \pi^2 n^2}{\gym^2} \ri)\sum_{\ell = 0}^{4n - 1} \frac{(-1)^{4n - \ell - 1}}{\G(4n - \ell)\ell!} \, f_{n,\ell}\, \gym^{-2(4n-1-\ell)}
    \\
    &= \Z^{k = 0}\,.
\end{aligned}
\right\}
\begin{aligned}
\,\, g_{2d} &\in -\mi \mathfrak{D}\,,
\\
g_{2d} &= -\mi \, \frac{\gym}{2}\,.
\end{aligned}
}
\eeq
\paragraph{The role of error functions} We noted in sec. \ref{sec:Resurgence_Higgs} that the 4d expression could equivalently be obtained from the non-perturbative data of an infinite series of error functions. This was part of our motivation to suspect the existence of a 2d interpretation of our sum over residues, following observations made in \cite{Witten:1992xu} on the role of error functions in the weak coupling expansion of 2d Yang-Mills on an arbitrary Riemann surface. We would like to comment on the relationship between the two-dimensional theory we have obtained and the error functions seen in sec. \ref{sec:Resurgence_Higgs}. 

It was shown explicitly in \cite{Griguolo:2024ecw} that the non-abelian localization of 2d Yang-Mills on an arbitrary Riemann surface can yield error functions by making use of the \textit{Heitler function}. In a distributional sense, the Heitler function defines integrals of the following kind
\beq \label{eq:Heitler_function}
    \int_0^\infty \diff x \, \text{exp}\le(\mp 2\pi \mi \, p \, x\ri) = \pm \frac{1}{2 \pi \mi} \text{pv} \frac{1}{p} + \frac{\delta(p)}{2}\,.
\eeq
In particular, it is the principal value term which leads to the perturbative component of the error function. The authors of \cite{Griguolo:2024ecw} showed that a product of integrals of this type yields derivatives of the right-hand side of \eqref{eq:Heitler_function} with respect to $p$. Results of this type have also been used in \cite{Leeb-Lundberg:2023jsj, Leeb-Lundberg:2025baz}. In our case, we have a similar result, except for the fact that our results yield only derivatives of the Dirac-delta function in \eqref{eq:Heitler_function}. Expressing
\beq
    \int_{-\infty}^0 \diff x \, \text{exp}\le(2\pi \mi \, p \, x \ri) + \int_0^\infty \diff x \, \text{exp}\le(2\pi \mi \, p \, x \ri)  = \d(p)\,,
\eeq
we note a clear analogy with the results of sec. \ref{sec:Resurgence_Higgs}, in particular the equations \eqref{eq:integral_splitting} and \eqref{eq:Z_discontinuity} showing cancellation between perturbative contributions. From the perspective of localization, we conjecture it is for this reason that our results extract only the non-perturbative data from the error functions. Thus, our intuition for the role of error functions found through resurgent analysis appears to be well-founded, and there exists a coherent picture of their role in the construction of the 4d and 2d path integrals.

\section{Discussion} \label{sec:discussion}
In this work, we have explored the physical interpretation of poles found in the localized expression for super-Yang-Mills theory on $\mathbb{S}^4$ for SU$(2)$ gauge group, coupled $N_f = 2 N_c = 4$ fundamental hypermultiplets, making the theory superconformal. In particular, we have deduced, by careful resurgent analysis, that the Coulomb branch integral is equal to the sum over the residues associated to these poles, \textit{only} through analytic continuation. We have given both a 4d perspective and a 2d perspective of the physics. From the 4d point-of-view, we have identified the poles with Seiberg-Witten monopoles arising from a Higgs branch localization scheme. From the 2d point-of-view, the poles are identified with unstable instantons of a theory with (0,2) supersymmetry. We have shown that these perspectives are compatible by reproducing the partition function of the 4d model, without 4d instanton contributions, with the 2d instantons playing the role of the mysterious non-perturbative objects in this sector. We have further found that such a 2d theory exists as a well-defined quantum theory precisely when the 4d theory is superconformal.

We have not discussed 4d instantons, nor insertions of the constituents of the chiral algebra components which, while compatible with the supersymmetry of the model, do not seem to contribute classically and would presumably affect the detailed form of the regularized one-loop determinants. It would be very interesting to see what modifications are needed in order to take these matters into account. Specifically, the authors of \cite{Pan:2017zie} conjectured that the two modifications should be related via the results of \cite{nekrasov2003seibergwittentheoryrandompartitions}.  

We also note that the authors of \cite{Beem_2015} use a relationship to a 2d $\left(0,4\right)$ supersymmetry algebra at intermediate stages in their derivation. It is not the case that this superalgebra is a symmetry of the chiral algebra theory. Presumably, the relevant relationship, at least in our context, is that the chiral algebra can be thought of as appearing from a twist, analogous to the A/2-twist, of a theory with $\left(0,4\right)$ supersymmetry. The resulting theory, restricted to operators in the supercharge preserved by the twist, can be shown to be conformal using arguments analogous to those of \cite{Beem_2015}. However, as with many twisted theories, it is not unitary.

We can demonstrate that such a theory is indeed compatible with our proposed matter content. The field content of our model is that of $N_{f}$ $\left(0,4\right)$ fundamental hypermultiplets, a $\left(0,4\right)$ vector multiplet, a $\left(0,4\right)$ adjoint hypermultiplet, and a $\left(0,4\right)$ adjoint Fermi multiplet. The regularization prescription for the partition function of the $\left(0,2\right)$ theory, introduced above,
is somewhat ad hoc. It may be the case that the $\left(0,4\right)$ version provides a justification for its use, similar to the justification for the use of the JK residue in some $\left(2,2\right)$ and $\left(0,2\right)$ computations. We hope to return to these matters using more concrete tools in the future.

Under the hypothesis described here, the poles in the integrand of the localized expression for the partition function of a superconformal $\mathcal{N}=2$ theory correspond to unstable instantons in an embedded 2d gauge theory. This is true in the Coulomb-branch version of the localization calculation on the four sphere. As we discussed in sec. \ref{sec:Higgs}, an alternative localization scheme, called Higgs-branch localization, can also be applied to this class of theories \cite{PanPeelaers}. Higgs-branch localization can also be applied to models in 2d \cite{Benini_2014_2D} where they lead to localization onto vortex configurations. The size of these configurations is controlled by a parameter in the localizing action, and Coulomb-branch localization is recovered when this parameter vanishes. It seems plausible that the relationship between the 4d Higgs-branch localization, whose results we have tied to the residues, and our 2d model goes through a similar 2d Higgs-branch localization. It would also be interesting to perform the four-dimensional Higgs-branch localization for our gauge group, to see how this compares with our results and to study the relationship between the Higgs-branch chiral ring and the chiral algebra chiral ring, as discussed in \cite{Beem_2015, Pan:2017zie}.

Our results extend beyond the SU$(2)$ gauge group to all superconformal gauge theories in four dimensions. Specifically, we have shown that these theories define 2d models that are free of gauge anomalies. It would be interesting to apply this to the study of quiver gauge theories, holographic models, (0,4) theories, or deformations of superconformal theories such as $\mathcal{N}= 2^*$ sYM. We hope to explore some of these relationships in future work.

Finally, it would be insightful to see to what extent resurgent structures can be extracted from these results. In particular, we can always take poles found in localized expressions as signalling the presence of error functions, and we can ask when the expressions are perturbative and thus genuinely resurgent? Another question worth pursuing would be whether a symmetry breaking deformation can be introduced to recover asymptotic behaviour \cite{Kozcaz:2016wvy,Dorigoni:2017smz} in perturbative calculations, for instance, the 1/2-BPS Wilson loop computed in \cite{Pestun:2007rz}.

\acknowledgments

This work was supported by the UKRI STFC consolidated grant ST/X000583/1, “New Frontiers in Particle Physics, Cosmology and Gravity”. I.A. was partially supported by UKRI EPSRC Early Career Fellowship EP/S004076/1.  J.R would like to thank A. Ferrari, M. Lemos, and A. Ratcliffe for useful discussions. I.Y would like to thank E. Lundberg and R. Panerai for helpful correspondence. J.R. is grateful to the Isaac Newton Institute for Mathematical Sciences, Cambridge, for support and hospitality during the programme ``Quantum field theory with boundaries, impurities, and defects'', where work on this paper was undertaken.

\appendix

\section{Residues}
\label{app:Residues}
In this appendix, we derive an explicit form for the residues of the zeroth instanton sector of $\Z$. Explicitly, expressing the Barnes--G functions in \eqref{eq:Z_SU(2)} as their product representations
\begin{equation}\label{eq:app_Z}
    \frac{\gym^3}{128\pi^{5/2}} \Z = \int^\infty_{-\infty}da\; a^2\; \text{exp}\left(\frac{-16\pi^2}{g^2_{\text{YM}}}a^2\right)\prod_{n=1}^\infty\frac{\left(1+\frac{4a^2}{n^2}\right)^{2n}}{\left(1+\frac{a^2}{n^2}\right)^{8n}}|Z_{\text{inst.}}|^2.
\end{equation}
To proceed, we exponentiate the integrand in the zeroth instanton sector
\begin{equation}\label{exponential-form}
\begin{aligned}
    \mathcal{I}(a)&=\text{exp}\left[-\frac{16\pi^2}{g^2_{\text{YM}}}a^2+2\text{ log(a)}+\sum_{n=0}^\infty\left(2n\text{ log}\le(1+\frac{4a^2}{n^2}\ri)-8n\text{ log}\le(1+\frac{a^2}{n^2}\ri)\right)\right]\\
    &=\text{exp}[f(a)].
\end{aligned}
\end{equation}
From the exponential form \eqref{exponential-form} we see that the poles are realised by $-$log$(0)$ terms. However, it also seems that there are divergences stemming from the infinite sum of terms like $\sim n\text{log}(n)$. We will see in fact these divergences mutually cancel, leaving a finite term which can be used to evaluate the full residues. This cancellation arises because we have chosen the superconformal matter content. The divergent logarithm term corresponding to the pole is removed as we shall see, by multiplying by $(a\pm \mi k)^{4k}$. The remaining components of the sums are
\beq\label{separated-sums}
\begin{aligned}
    (a\pm \mi k)^{4k}\cI(a)|_{a=\mp \mi k} =&\sum_{n=0}^\infty\le(2n~\text{log}\le(1+\frac{2k}{n}\ri)-8n~\text{log}\le(1+\frac{k}{n}\ri)\ri)
    \\
    +&\sum_{n=0}^{2k-1}2n~\text{log}\le(1-\frac{2k}{n}\ri)+\sum_{n=2k+1}^\infty2n~\text{log}\le(1-\frac{2k}{n}\ri)
    \\
    -&\sum_{n=0}^{k-1}8n~\text{log}\le(1-\frac{k}{n}\ri)-\sum_{n=k+1}^\infty8n~\text{log}\le(1-\frac{k}{n}\ri)
    \\
    +&8k~\text{log}(k)-4k~\text{log}(2k)+4k~\text{log}(2),
\end{aligned}
\eeq
where the first two terms in the final line come from the terms in the divergent logarithms that are not cancelled, and the third term comes from normalising the pole at $n=2k$. By shuffling around the various sums, we can reduce \eqref{separated-sums} down to
\beq
    (a\pm \mi k)^{4k}\cI(a)|_{a=\mp \mi k} =\sum_{n=1}^{k}(12n-8k)\text{log}(n)-\sum_{n=k+1}^{2k}(4n-8k)\text{log}(n)+2k\pi i.
\eeq
We can conclude that for a pole at $a=\pm i k$, the exponential evaluates to
\beq
    \label{eq:def-kappa}
    \kappa(k)\coloneqq -2^{4k}k^2 \, \text{exp}\le[\frac{16\pi^2k^2}{g_{\text{YM}}}+\;\;\sum_{n=1}^{k}(12n-8k)\text{log}(n)
    -\sum_{n=k+1}^{2k}(4n-8k)\text{log}(n)\ri],
\eeq
where we drop the $2k\pi i$ term since the exponential of this is unity. 

To determine the residue of the pole of order $4k$ we evaluate $4k-1$ derivatives of the following quantity
\beq
    \mathcal{J}(a)=(a \mi \pm k)^{4k}\cI(a)\big|_{a=\pm \mi k} = u^{4k}\cI(u)|_{u = 0}, 
\eeq
where we have introduced the variable $u = a\mi \pm k$ to shift all poles to the origin, and re-write them in a universal form. This removes the pole singularities and is regular in a neighbourhood of $u = 0$. Thus the leading term in the $(4k-1)^{th}$ derivative will be the residue of the pole at $\pm \mi k$. To determine $\mathcal{J}^{(p)}(u)$ we start by again exponentiating
\beq
    \mathcal{J}(u) = \text{exp}\le[ 4k \,\text{log}(u) + f(u) \ri] = \text{exp}\le[ f_{\text{reg.}}(u) \ri].
\eeq
where $f_{\text{reg.}}(u)$ is the regular component of $f(u)$ ($4k$log$(u)$ regularizes the singular part of $f(u)$ as noted above). The $p^{th}$ derivative is then given generically by Faà di Bruno's formula
\beq
    \frac{d^p}{dx^p} f(g(x)) = \sum^p_{l=0} f^{(p)}(g(x)) \cdot B_{p,l}\le( g^\prime(x), g^{\prime \prime}(x), \hdots, g^{(p - l + 1)}(x)\ri),
\eeq
where $B_{p,l}(x_1,\hdots,x_{p - l + 1})$ are Bell polynomials. In our case we identify
\beq
    f(g(u)) = \mathcal{J}(u) = \text{exp}(g(u)), \quad g(u) = f_{\text{reg.}}(u).
\eeq
Furthermore, as $\mathcal{J}^{(p)}(u) = \mathcal{J}(u)$ we have
\beq \label{eq:Faa_reg}
\begin{aligned}
    \mathcal{J}^{(p)}(u) 
    & = \mathcal{J}(u)\cdot B_p\le( f_{\text{reg.}}^\prime(u), f_{\text{reg.}}^{\prime \prime}(u),\hdots, f_{\text{reg.}}^{(p)}(u) \ri)
\end{aligned}
\eeq
where $B_p$ are now the complete Bell polynomials. We are then tasked with computing $f^{(i)}_{\text{reg.}}(u)$, 
 Starting with the first derivative,
\beq \label{eq:first_derivative}
\begin{aligned}
    \frac{d f(u)}{du} =  & \, \mi \frac{32 \pi^2}{\gym^2}(u + k) + \frac{2 \mi}{u + k}
    \\
    &- 8\mi (u + k) \le[ \zeta_1(2(u + k)) + \zeta_1( -2(u + k)) - \zeta_1(-(u + k)) - \zeta_1(u + k)\ri],
\end{aligned}
\eeq
where $\zeta_1$ are Hurwitz zeta functions defined as,
\beq
    \zeta(s,a)=\sum^\infty_{n=0}\frac{1}{(n+a)^s};\quad\zeta(1,a)\coloneqq\zeta_1(a).
\eeq
While the Hurwitz zeta function has a pole for $s=1$ the combination seen in \eqref{eq:first_derivative} does not. To see this, note that the Laurent series for $\zeta(s,a)$ is
\begin{equation}
    \zeta(s,a)=\frac{1}{s-1}+\sum_{n=0}^\infty\frac{(-1)^n}{n!}\gamma_n(a)(s-1)^n.
\end{equation}
We regulate the sums of zeta functions in \eqref{eq:first_derivative} by taking $s=1+\epsilon$ for $\epsilon \ll 1$.
In the limit $\epsilon\to0$, this evaluates to
\begin{equation}\label{regularised_zetas}
\begin{aligned}
    \psi(u+k)+\psi(-(u+k))-\psi(2(u+k))-\psi(-2(u+k)) \equiv \psi_k+\psi_{-k}-\psi_{2k}-\psi_{-2k},
\end{aligned}
\end{equation}
where $\gamma_0(z)=-\psi(z)$ and $\psi(z)=\Gamma'(z)/\Gamma(z)$ is the digamma function. We finally find for \eqref{eq:first_derivative}:
\beq \label{eq:reduced_d1}
\begin{aligned}
    f^\prime(u) = \mi \frac{32 \pi^2}{\gym^2}(u + k) + \frac{2 \mi}{u + k} - 8\mi (u + k) \le[ \psi_k+\psi_{-k}-\psi_{2k}-\psi_{-2k}\ri].
\end{aligned}
\eeq
Higher order derivatives of $f_{\text{reg.}}(u)$ at $u = 0$ will correspond to coefficients in the Laurent series of \eqref{eq:reduced_d1}:
\beq
    f^\prime(u) = \frac{b_{k,-1}}{u} + \sum_{n = 0 }^\infty b_{k,n} u^n\,,
\eeq
where the indices $k, n$ refer to the position of the pole and the order in the expansion, respectively. For instance, $f^{\prime \prime}_{\text{reg}}(0) = 2 b_{k,1}$. Using the following expansions of the digamma functions
\beq
\begin{aligned}
    \psi_k + \psi_{-k} - \psi_{2k} - \psi_{-2k} = & \, \frac{1}{2u} + \sum_{n=0}^\infty \le[2 (2^n - 1)\le(\zeta(n+1)\d_{n\text{ mod } 2,0} - (-1)^n H_{k,n+1}\ri)\ri.
    \\
    &\le. \phantom{\frac{1}{2u} +} + \frac{3}{2}\frac{1}{(-k)^{n+1}} + \sum_{j=1}^{k - 1}\frac{(-2)^{n+1}}{(2k - j)^{n+1}}\ri]u^n
    \\
    \equiv & \,\frac{1}{2u} + \sum_{n = 0}^\infty a_{k,n} u^n\,,
\end{aligned}
\eeq
where $H_{a,b}$ is the $a^{th}$ generalised harmonic number of order $b$, defined as
\beq
    H_{a,b} = \sum_{j = 1}^{a}\frac{1}{j^b},
\eeq
and we employ an abuse of notation that $\zeta(1)=\g$, Euler's constant, we find that
\beq
\begin{aligned}
    f^\prime(u)=&~\le(\frac{b_{-1}}{u}+b_0+b_1u+\sum_{n=2}^\infty b_nu^n\ri)\,.
\end{aligned}
\eeq
In the above equation, the coefficients $b_n$ are given by
\beq
\begin{aligned}
    b_{-1}&=-4\mi k
    \\
    b_0&=\mi\frac{32\pi^2}{g_{\text{YM}}^2}k+\sum_{j=k+1}^{2k}\frac{16 \mi \,k}{j}+\frac{4\mi}{2k}
    \\
    b_1&=\mi\frac{32\pi^2}{g_{\text{YM}}^2}-8\mi \,ka_1-2a_0-\frac{4\mi}{2k^2}
    \\
    b_n&=-8\mi \, ka_n-8 \mi\, a_{n-1}+2\mi \frac{(-1)^n}{k^{n+1}},~n>1.
\end{aligned}
\eeq
We now have all the data required to compute the residues. To do so, we must evaluate \eqref{eq:Faa_reg} at $u = 0$,
\beq \label{eq:appendix_residues}
    \text{Res}[u = 0,k] = 128 \pi^{5/2} \frac{\k(k)}{(4k - 1)!}\cdot B_{4k-1} (0!b_0, 1!b_1, \hdots (4k-1)!b_{4k-1})\,,
\eeq
where $\kappa(k)$ is defined in \eqref{eq:def-kappa}.
The remaining coefficients are then
\beq \label{eq:f_hat}
    \hat{f}_{k,l}(\gym)= 128 \pi^{5/2} \,\frac{\kappa(k)}{l!}\cdot B_l(0! \, b_{k,0},1!\,b_{k,1},\hdots,l!\,b_{k,l}).
\eeq
We note that these coefficients are for the full Laurent expansion of \eqref{eq:I(a)} and should be distinguished from the coefficients appearing in \eqref{eq:series_I_no_exp}, which are the Laurent series coefficients for the integrand excluding the exponential component. The latter can be obtained from the former by taking $\gym \to \infty$, which is equivalent to setting the exponential in \eqref{eq:app_Z} to $1$. This limit of the coefficients we will denote by $\hat{f}_{k,\ell}$, with no $\gym$--dependence.

\paragraph{Conversion to Laplace variable}
For the purposes of writing $\Z$ in terms of error functions, we require the coefficients from the Laurent series of $\cI$ in terms of the Borel variable, $s$. In this variable, the result is
\beq \label{eq:fnl}
\begin{aligned}
    f_{k,l}(\gym)=-\mi\frac{2(32\pi^2k)^{4k-1-l}}{k^{l}}&\sum_{n=0}^l\le(-\frac{1}{2}\ri)^nB_{l,n}\le(-1!!,1!!,3!!,\hdots,(2(l-n)-1)!!\ri)
    \\
    &\times\le(\sum_{j=0}^n(-2)^j(n)_j(4k)_{n-j}\sum_{m=0}^j(-ik)^m\hat{f}_{k,m}(\gym)\ri)
\end{aligned}
\eeq
where $(i)_j$ is the falling factorial $i!/(i - j)!$ and the coefficients $\hat{f}_{k,l}(\gym)$ were defined in \eqref{eq:f_hat}. As before, the coefficients in \eqref{eq:appendix_residues} can be obtained from \eqref{eq:fnl} by taking $\gym \to \infty$. We again denote the coefficients in this limit by $f_{k,\ell}$. 

For $\ell=0$, the expression \eqref{eq:fnl} simplifies to
\beq
    f_{k,0}=\mi\cdot k^{1+4k}2^{3+24k}\pi^{8k+1/2}\text{exp}\le[\sum_{n=1}^{k}(12n-8k)\text{log}(n)-\sum_{n=k+1}^{2k}(4n-8k)\text{log}(n)\ri].
\eeq
These can also be readily seen to agree with table 2 in \cite{Aniceto:2014hoa}.

\bibliographystyle{JHEP}
\bibliography{references}

\end{document}